\documentclass[aps,showpacs,superscriptaddress,twocolumn,8pt,nofootinbib]{revtex4-1}

\usepackage{dcolumn}
\usepackage{graphicx}  
\usepackage{dcolumn}   
\usepackage{bm}        
\usepackage{amsthm,amssymb}   
\usepackage{amsmath}
\usepackage{dsfont}
\usepackage{xcolor}
\usepackage{mathrsfs}
\usepackage{amsfonts}
\usepackage{amsthm}
\usepackage{mathrsfs}
\usepackage{cases}
\usepackage{url}
\usepackage{framed}
\usepackage{cancel,soul,ulem}
\usepackage{mathtools}
\usepackage{graphicx,psfrag,subfigure}
\usepackage{multirow}
\usepackage{gensymb} 
\usepackage{hyperref}

\newcommand{\BEQ}{\begin{equation}}
\newcommand{\EEQ}{\end{equation}}
\newcommand{\bea}{\begin{eqnarray}}
\newcommand{\eea}{\end{eqnarray}}
\newcommand{\be}{\begin{equation}}
\newcommand{\ee}{\end{equation}}
\newcommand{\BEA}{\begin{eqnarray}}
\newcommand{\EEA}{\end{eqnarray}}

\def\:={\,\raisebox{0.85pt}{.}\hspace{-2.78pt}\raisebox{2.85pt}{.}\!\!=\,}
\def\=:{\,=\!\!\raisebox{0.85pt}{.}\hspace{-2.78pt}\raisebox{2.85pt}{.}\,}

\begin{document}

\title{
Using random testing in a feedback-control loop to manage a safe exit
from the COVID-19 lockdown
      }

\author{Markus M\"uller}
\email{Markus.Mueller@psi.ch}
\affiliation{
Paul Scherrer Institut, CH-5232 Villigen PSI, Switzerland}
\author{Peter M. Derlet}
\affiliation{
Paul Scherrer Institut, CH-5232 Villigen PSI, Switzerland}
\affiliation{
Department of Materials, ETH Zurich, CH-8093 Zurich, Switzerland}
\author{Christopher Mudry}
\affiliation{
Paul Scherrer Institut, CH-5232 Villigen PSI, Switzerland}
\affiliation{Institut de Physique, EPF Lausanne, Lausanne, CH-1015, Switzerland}
\author{Gabriel Aeppli}
\affiliation{
Paul Scherrer Institut, CH-5232 Villigen PSI, Switzerland}
\affiliation{Institut de Physique, EPF Lausanne, Lausanne, CH-1015, Switzerland}
\affiliation{Department of Physics, ETH Zurich, Zurich, CH-8093, Switzerland}
\date{\today}

\begin{abstract}

We argue that frequent sampling of the fraction of infected people
(either by random testing or by analysis of sewage water),
is central to managing the COVID-19 pandemic
because it both measures in real time
the key variable controlled by restrictive measures, and anticipates
the load on the healthcare system due to progression of the
disease. Knowledge of random testing outcomes will (i) significantly
improve the predictability of the pandemic, (ii) allow informed and
optimized decisions on how to modify restrictive measures, with much
shorter delay times than the present ones, and (iii) enable the
real-time assessment of the efficiency of new means to reduce
transmission rates.

Here we suggest, irrespective of the size of a suitably homogeneous
population, a conservative estimate of 15'000 for the number of
randomly tested people per day which will suffice to obtain reliable
data about the current fraction of infections and its evolution in
time, thus enabling close to real-time assessment of the quantitative
effect of restrictive measures.
Still higher testing capacity permits detection of geographical
differences in spreading rates. Furthermore and most importantly, with
daily sampling in place, a reboot could be attempted while the
fraction of infected people is still an order of magnitude higher than
the level required for a relaxation of restrictions with testing focused on
symptomatic individuals. This is
demonstrated by considering a feedback and control model of mitigation
where the feed-back is derived from noisy sampling data.

\end{abstract}

\maketitle


\section{Introduction}
\label{sec: Introduction}

The COVID-19 pandemic has led to a worldwide shutdown of
a major part of our economic and social activities. This political measure
was strongly suggested by epidemiologic studies
assessing the cost in human lives depending on different possible
policies (doing nothing, mitigation, suppression)%
~\cite{Ferguson06,Imperial College,COVID-19REPORT09,FergusonReport13}.
Mitigation can be achieved by combinations of different measures, including 
physical distancing, contact tracing, restricting public gatherings,
and the closing of schools, but also the testing for infections.
The quantitative impact of very frequent testing of the entire population
for infectiousness has been studied in a recent unpublished work
by Jenny et al.\ in Ref.\
\cite{Jennyreport20}. We will estimate in Sec.\ \ref{sec: Massive testing}
that to fully suppress the COVID-19 pandemic by widespread testing for
infections, one needs a capacity to test millions of people per day
in Switzerland. This should be compared to the present number
of 7'000 tests per day across Switzerland.%
~\footnote{
As of early April 2020, according to the liveticker of
the Swiss radio and television
\url{https://www.srf.ch/}.
          }
Here we suggest that by testing a much smaller number of
randomly selected people per day one can obtain important quantitative
information on the rates of transmission, so as to enable well-informed
decisions.

\begin{figure}[t]
\begin{center}
\includegraphics[width=0.95\linewidth]{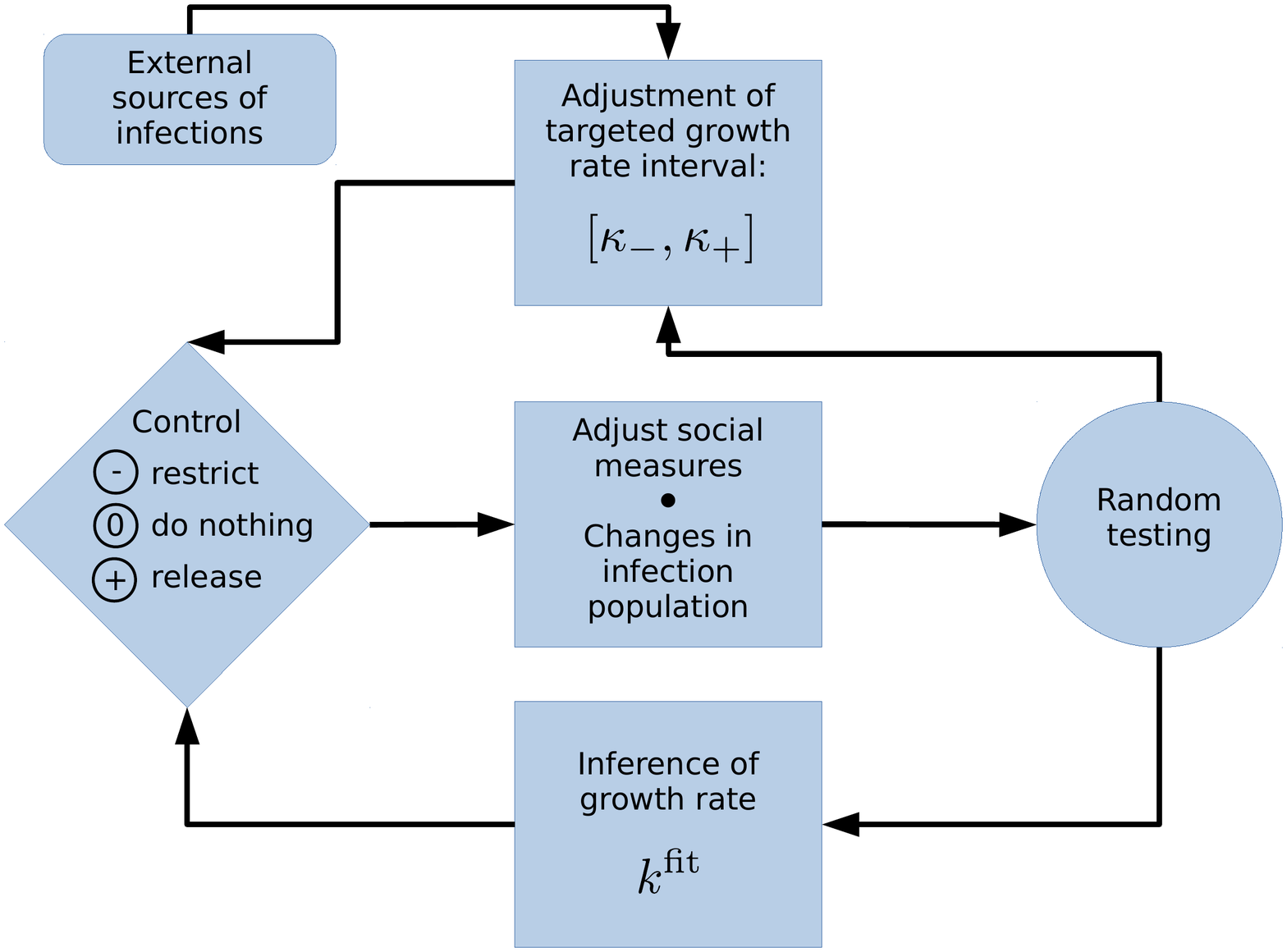}
\end{center}
\caption{
Feedback and control loop that allows stabilization of the pandemic.
The key quantity measured by random testing is the growth rate $k$ of
infection numbers.  If $k$ exceeds a tolerable upper threshold
$\kappa^{\,}_{+}$, restrictions are imposed. For $k$ below a lower threshold
$\kappa^{\,}_{-}$, and if infection numbers are below critical, restrictions
are released. In the absence of a substantial influx of infected
people from outside the country, and provided infection numbers are
below a critical value, the optimal target of the growth rate is $k=0$,
corresponding to a marginally stable state, where infections neither
grow nor decrease exponentially with time. If higher testing rates are
available, the measured observables and control strategies can be
geographically refined, particularly to avoid hotspots.
        }
\label{fig: concept}
\end{figure}

Figure \ref{fig: concept}
summarizes the key concept of the paper, namely a feedback and control
model for the pandemic. The essential output from random testing is the
growth rate of the number of currently infected people, which itself is
regulated by measures such as those enforcing physical distances
between persons (physical distancing),%
~\footnote{
We prefer the terminology
\textit{physical distancing} to \textit{social distancing}.  
          }
and whose tolerable values are fixed by the capacity of the health-care
system. A feedback and control approach%
\cite{Wiener48},
familiar from everyday
implementations such as thermostats regulating heaters and air
conditioners, should allow policy makers to damp out oscillations in
disease incidence which could lead to peaks in stress on the
health-care system as well as the wider economy.
Any other measurement of the fraction of currently infected people can
replace the random testing, for example there are proposals
to estimate this fraction from analysis of sewage water with PCR tests%
~\cite{Medema20,Wu20}.

\begin{figure}[t]
\begin{center}
\includegraphics[width=0.95\linewidth]{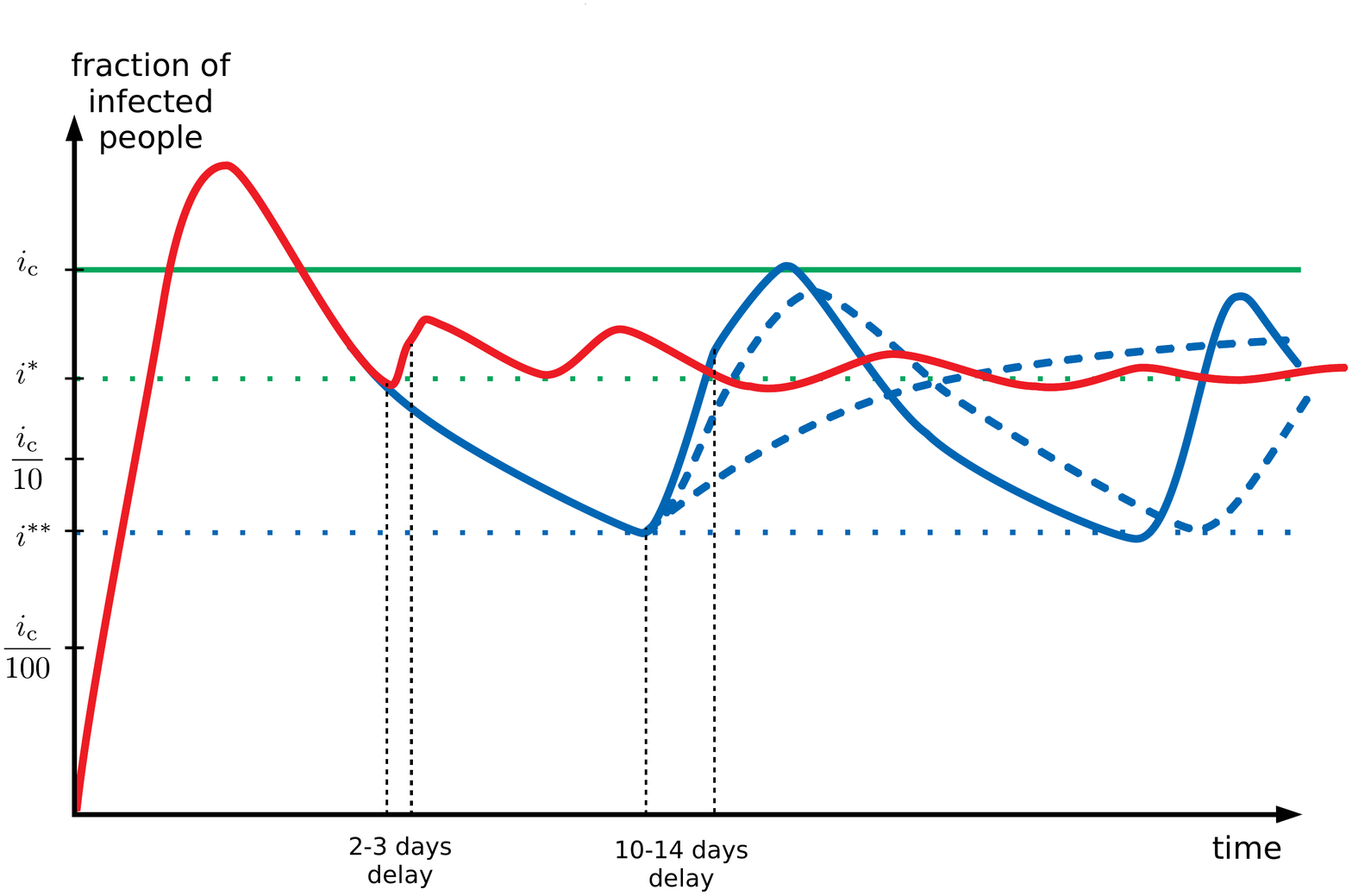}
\end{center}
\caption{
Dynamics of the pandemic with and without a feedback and control
scheme in place, as measured by the fraction $i$ of currently infected people
(logarithmic scale).  After the limit of the health system,
$i^{\,}_{\mathrm{c}}$, has been reached, a lockdown brings $i$
down again. The exponential rate of decrease is expected to be very
slow, unless extreme measures are imposed. The release of measures
upon a reboot is likely to re-induce exponential growth,
but with a rate difficult to predict.
Three possible outcomes are shown in blue curves in
the scenario without testing feedback, where the effect of the new
measures becomes visible only after a delay of 10-14 days.
In the worst case, $i$ grows by a multiplicative factor of order 20
before the growth is detected. A reboot can thus be risked only once
$i\leq i^{**}\equiv i^{\,}_{\mathrm{c}}/20$,
implying a very long time in lockdown after the initial peak.
Due to the long delay until policy changes show observable effects,
the fluctuations of $i$ will be large.
Random testing (the red curve) has a major advantage. It measures $i$
instantaneously and detects its growth rate within few days, whereby the
higher the testing rate the faster the detection. Policy adjustments can thus
be made much faster, with smaller oscillations of $i$. A safe reboot
is then possible much earlier, at the level of
$i\leq i^{*}\approx i^{\,}_{\mathrm{c}}/4$.
        }
\label{fig: comparison}
\end{figure}

An important further benefit of our feedback and control scheme is
that it allows a much faster and safer reboot of the economy than with
feedback through confirmed infection numbers of symptomatic persons or
deaths \cite{FergusonReport13},
for the latter, secondary indicator is heavily delayed and reflects
the state of the pandemic only incompletely as it does not account
for asymptomatic carriers. Figure \ref{fig: comparison}
illustrates the resulting difference in the ability to control the disease.

Without feedback and control informed by a primary indicator,
analogous to the temperature provided by the thermometer
in the thermostat example,
measurable in (near) real time,
there is a huge lapse between policy changes and the
observable changes in numbers of positively tested people.
To relax restrictions safely, the fraction of currently infected
people must decrease to a level $i^{**}$ such that a subsequent
undetected growth during 10-14 days will not move it above
the critical fraction $i^{\,}_{\mathrm{c}}$ manageable by the
health-care system.  The current situation where we are mainly looking
at lagging secondary indicators, namely infection rates among symptomatic
individuals or even deaths, is comparable to driving a car from
the back seat and with knowledge only of the damage caused
by previous collisions.
To minimize harm to the occupants of the vehicle, driving very
slowly is essential, and oscillations from a straight course
are likely to be large. 

Daily random testing reduces the delay between changes in
policy and the observation of their effects very
significantly. Moreover, it directly measures the key quantity of
interest, namely the fraction of currently infected people and its growth rate,
information that is very valuable to gauge further interventions. Such
information is much harder to infer from data about positively tested
patients only, by fitting it to specific epidemiological models with
their inherent uncertainties. The shortened time delay due to
feedback and control allows a reboot to be attempted at much higher
levels of infections, $i^{*}>i^{**}$, which implies a much shorter
time in lockdown.

We point out before proceeding further that this is a contribution
from physicists that makes simplifying assumptions inconsistent with
details of medical and epidemiological reality to obtain some key
estimates and illustrate the basic principles of feedback and control
as applied to the current pandemics. When reduced to practice, special
attention will need to be paid to all aspects of the testing
methodology, from the underlying molecular engineering paradigm
(e.g., PCR) and associated cost/performance trade offs, to population
sample selection consistent with societal norms and statistical needs,
and safe (i.e., not risking further infections) operation of testing
sites. Furthermore, in preparation for the day when more is
known about the immune response to COVID-19 and possible vaccines, we
plan to revise our models for feedback derived from a
reliable immunoassay with well-specified performance parameters, such as lag
times with respect to infection. 

The paper is organized as follows.
We summarize and explain the key findings in simple terms
in Sec.\ \ref{sec: Summary}.
In Sec.\ \ref{sec: Massive testing}, we discuss the use of
massive testing as a direct means to contain the pandemics, showing
that it requires a 100-fold increase of the current testing frequency.
In Sec.\ \ref{sec: Quantifying the effectiveness of restrictions},
we define the main challenge to be addressed:
To measure the quantitative
effect of restrictive measures on the transmission rate.
Section \ref{sec: Smart statistical testing}
introduces the idea of randomized testing.
Section \ref{sec: National modeling and intervention}
constitutes the central part of the paper, showing
how data from sparse sampling tests can be
used to infer essentially instantaneous growth rates,
and their regional dependence. We define a model of policy interventions 
informed by feed-back from random testing and analyze it theoretically.
The model is also analyzed numerically in Sec.\,
\ref{sec: Simulation of mitigation ...}.
In Sec.\ \ref{sec: Regionally refined reboot  and mitigation strategies},
we generalize the model for regionally refined analysis of the
epidemic growth pattern which becomes the preferred choice if higher
testing rates become available.
We conclude with Sec.\ \ref{sec: Summary and conclusion} by
summarizing our results and their implication for a safe reboot after
the current lockdown.  In the Supplementary Information,
we address contact
tracing and argue that it can complement, but not substitute for
random testing. Finally, we present the algorithm used for our numerical
results.

\section{Summary of key results}
\label{sec: Summary}

We argue that the moderate number of 15'000 random tests per day
yields valuable information on the dynamics of the disease.
Assuming that at a given time a conservatively estimated fraction of about
$i^{*}\approx 0.07\%$
of the population is currently infected [see Eq.\ \ref{eq: estimate i*}],
on the order of 10 infected people will be detected every day.
Can such a small number of detected
infections be useful at all, given that these numbers fluctuate
significantly from day to day? The answer is yes. We show that after a
few days the acquired signal becomes stronger than the noise level.
It is then possible to establish
whether the infection number is growing or decreasing and,
moreover, to obtain a quantitative estimate of the instantaneous
growth rate $k(t)$.

One of our central results is Eq.~(\ref{Deltat1}) for the time where
the signal becomes clear, which we rewrite in the simplified form
\be
\Delta t^{\,}_{1}=
\frac{C}{(k^{2}_{1}\,r)^{1/3}},
\label{Deltat1simple}
\ee
where $k^{\,}_{1}$ is the current growth rate of infections to be detected,
and $r$ is the number of tests per day%
~\footnote{
If the fraction of infected people can be measured via sewage water,
$r$ will be related to the number of people connected
to a given sewage plant. But at this point, the relationship
between such data and the actual current number of infections remains
a topic for research. Of course, once key parameters such as the lag time
between infection and incidence of biomarkers in sewage are known,
sewage tests could become highly competitive.
          }.
The numerical constant $C$ depends on the required signal to noise ratio.
A typical value when
detecting large values of $k^{\,}_{1}$ is $C\approx 30-40$.

This result shows that the higher the number of tests $r$ per day, the
shorter the time to detect a growth or a decrease of the infected
population. The smaller the current growth rate $k^{\,}_{1}$,
the longer the time to detect it above the noise inherent to
the finite sampling.

How long would it take to detect that a release of restrictive
measures has resulted in a nearly unmitigated growth rate of the order
of $k^{\,}_{1} =0.23$ (which corresponds to doubling every 3 days)? Even with
a moderate number of $r=15'000$ per day, we find that within only
$\Delta t^{\,}_{1}\approx 3-4$ days such a strong growth will emerge above
the noise level, such that countermeasures can be taken
(see Fig.~\ref{fig: Delta time 1}). During this short time, the damage remains
limited. The infection numbers will have risen by a multiplicative
factor between $2$ and $3$. This degree of control must be compared to
a situation where no information on the current growth rate is
available, and where the first effects of a new policy are seen in the
increased number of symptomatic, sick people only 10-14 days
later. Over this time span, with a growth rate of $k^{\,}_{1}=0.23$, the
infection numbers will have grown by a factor of 10-30 before one
realizes eventually that an intervention must be made.

Random testing decreases both the time scale until informed policy
adjustments can be taken and the temporal fluctuations of the
infection numbers.  As in any feedback and control loop, the more
frequent the testing is, the shorter are the delay times, and thus the
smaller are the fluctuations. The various benefits of increasing the
testing frequency are shown in Fig.~\ref{fig: Performance fct r}, which are
obtained by simulating a specific mitigation strategy, where we built
in the uncertainty about the efficacy of political interventions. The
shorter delay times and the reduced fluctuations result in decreased
strain on the health system, lower economic costs, and a lower number
of required interventions.
 
In addition to these benefits, a higher testing rate $r$ also opens
the opportunity to analyze geographic differences and refine the
mitigation strategy accordingly, as we discuss in Sec.\
\ref{sec: Regionally refined reboot and mitigation strategies}. 
 
\section{Massive testing}
\label{sec: Massive testing}

If the massive frequency of 1.5 million tests per
day becomes available in Switzerland, it will be possible
to test any Swiss resident every 5 to 6 days.
If the infected people that have been
detected are kept in strict quarantine (such that they will not infect
anybody anymore with high probability), such massive testing could be
sufficient to prevent an exponential growth in the number of cumulated
infections without the need of draconian physical distancing
measures. We now explain qualitatively our approach to reach this conclusion
(Ref.~\cite{Jennyreport20} gives a more detailed quantitative analysis).

The required testing rate can be estimated as follows.
Let $\Delta T$ denote the average time until an infected person
infects somebody else. The reproduction number $R$, i.e.,
the number of new infections transmitted on average by an infected person,
falls below $1$ (and thus below the threshold for exponential growth)
if non-diagnosed people are tested at time intervals of no more than
$2\Delta T$.
Thus, the required number of tests over the time  $2\Delta T$,
the full testing rate $\tau^{-1}_{\mathrm{full}}$, is
\begin{subequations}
\be
\tau^{-1}_{\mathrm{full}}=
\frac{N^{\,}_{\mathrm{CH}}}{2\Delta T},
\label{minimalrate}
\ee
where
\be
N^{\,}_{\mathrm{CH}}=
8'500'000
\ee
\end{subequations}
is the number of inhabitants of Switzerland.%
~\footnote{
Note that if tests take
the nonvanishing time $t^{\,}_{\mathrm{test}}$ to yield a diagnosis,
this time needs to be subtracted from the denominator in
Eq.\ (\ref{minimalrate}),
thereby resulting in an {increase} of the full testing rate 
$\tau^{-1}_{\mathrm{full}}$.
          }
Without social restrictions, it is estimated that%
~\cite{Liu20}
\begin{subequations}
\be
\Delta T\approx
3\,\mathrm{days}, 
\ee
such that
\be
\tau^{-1}_{\mathrm{full}}=1.4\times10^{6}/\mathrm{days},
\ee
\end{subequations}
i.e., about
1.4 million tests  per day would be required
to control the pandemics by testing only.
If additional restrictions such as physical distancing etc., are imposed,
$\Delta T$  increases by a modest factor
and one can get by with indirectly proportionally
fewer tests per day. Nevertheless, on the order of 1 million tests per day
is a minimal requirement for massive testing to contain the pandemics
without further measures.

However, even while the Swiss capabilities are still far from reaching
1 million tests per day, testing for infections offers two important
benefits in addition to identifying people that need to be
quarantined. First, properly randomized testing allows to monitor and
study the efficiency of measures that keep the reproduction number $R$
below 1. This ensures that the growth rate $k$ of case numbers and new
infections is negative, $k<0$.  Second, frequent testing, even if
applied to randomly selected people, helps suppress the reproduction
number $R$ and thus allows policy to be less restrictive in terms of
other measures, such as physical distancing.

To quantify the latter benefit, observe that
the effect of massive testing on the growth rate $k$ is proportional
to the testing rate%
~\cite{Jennyreport20}.
Let us assume that without testing or social measures
one has a growth rate $k^{\,}_{0}$. Then,
if the testing rate $\tau^{-1}_{\mathrm{full}}$ is sufficient
to completely suppress the exponential growth in the absence of other
measures, a smaller testing rate $\tau^{-1}$ decreases the growth rate
$k^{\,}_{0}$ down to $(\tau^{-1}/\tau^{-1}_{\mathrm{full}})\times k^{\,}_{0}$.
The remaining reduction of $k$ to zero must then be achieved by
a combination of restrictive social measures and contact tracing.

It is possible to refine the argument above to take account of the
possibility of a spectrum of tests with particular cost/performance
trade offs, i.e., a cheaper test with more false negatives
could be used for random testing, whereas those displaying symptoms
would be subjected to a ``gold standard'' (PCR) assay of viral genetic
material.

\section{Quantifying the effectiveness of restrictions}
\label{sec: Quantifying the effectiveness of restrictions}

A central challenge for establishing reliable predictions
for the time evolution of a pandemic
is the quantification of the effect of social restrictions on the
transmission rate%
~\cite{COVID-19REPORT09}.
Policymakers and epidemiologists urgently need
to know by how much specific restrictive measures reduce the growth
rate $k$. Without that knowledge, it is essentially impossible to take
an informed decision on how to optimally combine such measures to
achieve a (marginally) stable situation, defined by the condition
of a vanishing growth rate 
\begin{equation}
k=0.
\label{eq: definition marginal stable state}  
\end{equation}
Indeed, marginal stability is optimal for two reasons.
First, it is sustainable in the sense that the burden
on the health system does not grow with time.
Second, it is the least economically and socially
restrictive state compatible with the stability requirement.

In Secs.\ \ref{sec: Smart statistical testing}
and \ref{sec: National modeling and intervention},
we suggest how marginal stability can be achieved,
while simultaneously measuring
the effects of a particular set of restrictions.

\section{ Statistical testing}
\label{sec: Smart statistical testing}

We claim that statistically randomized testing can be used in a
smart way, so as to keep the dynamics of the pandemics under
control as per the feedback loop of Fig.~\ref{fig: concept}.
We emphasize that this is possible without the current time
delays of up to 14 days. The latter arises since we only observe
confirmed infections stemming from a highly biased test group that
eventually shows symptoms long after the initial infection has
occurred.

The idea of statistical testing is to \textit{randomly} select people%
~\footnote{
It is important that the set of randomly selected people
must change constantly, so that it should happen extremely rarely that
a given person is tested twice.
          }  
and test them for infectiousness.%
~\footnote{
Here, we solely focus on a person being infectious,
but \textit{not} on whether the person has developed antibodies.
The latter test indicates that the person has been infected any time
in the past. Serological tests for antibodies and (potential) immunity have
their own virtue, but aim at different goals from the random
testing for infections that we advocate here.
By following the fraction of infections as a function of time, we can
determine nearly instantaneously the growth rate of infections, $k(t)$,
and thus assess \textit{and quantify} the effectiveness of
socio-economic restrictions through the observed changes in $k$
following a change in policy. This monitoring can even be carried out in a
regionally resolved way, such that subsequently,
restrictive or relaxing measures can be adapted to
different regions (urban/rural, regions with
different degrees of immunization, etc.).
          }

We stress that randomized testing is essential to obtain information on the
current number of infections and its evolution with time.  It serves
an additional and entirely different purpose from testing people with
symptoms, medical staff, or people close to somebody who has been
infected, all of whom constitute highly biased groups of people.

The first goal of random testing is to obtain a firm test/confirmation
of whether the current restrictive measures are sufficient to mitigate
or suppress the exponential growth of the COVID-19 pandemic,%
~\footnote{
A suppression of the COVID-19 pandemic is achieved if,
for a sufficiently long time, the number of infections
decays exponentially with time. Mitigation aims to reduce the
exponential rate of growth in the number of infections.
Stability is achieved when that number tends to a  constant.
Once stability is reached, one may start relaxing the restrictions
step by step and monitor the effect on the growth rate $k$ as a
function of geographic regions.
          }
and whether the effectiveness differs from region to region. In case the
measures should still be insufficient, one can measure the current
growth rates and monitor the effect of additional restrictive measures.

\section{National modeling and intervention}
\label{sec: National modeling and intervention}

We first analyze random testing for the case where we treat the
country as a single entity with a population $N$. This will allow
us to understand how testing frequency affects key characteristics of
policy strategies.

\subsection{Model assumptions}

We consider a model with the following idealizing assumptions:
\begin{enumerate}
\item[(1)]
An unbiased representative sample of the population is tested.
A bias may underestimate the most relevant growth rate. 
 
\item[(2)]
The rate of false positive tests is much less than the expected frequency
of detection of infections.

\item[(3)]
Tests show whether a person is acutely infected in a short time
(on the order of one day).

\item[(4)]
Policy measures can be applied rapidly, and their effect is immediate.
Time delays due to the adaptation of human behavior to new rules is neglected.

\item[(5)]
The population is homogeneous as far as interactions between its members
are concerned, e.g., there are no (semi)-isolated subpopulations.
We do not account for large deviations in infectiousness that may lead
to superspreading events%
~\cite{LloydSmith05}.

\end{enumerate}
As is well known to epidemiologists and the medical profession, the
assumptions (1-4) clearly are violated to varying degrees in reality,
but they can be taken into account by refinements of our model, whose
operating principles and basic behavior will remain qualitatively the
same. On the other hand, violations of assumption (5) can lead to new
and dangerous effects, namely hotspots related to Anderson
localization~\cite{Anderson58}, which we discuss in Sec.\
\ref{sec: Regionally refined reboot  and mitigation strategies}.

Let $U$ be the actual number of currently infected,
but yet undetected people.
(As in Ref.~\cite{Jennyreport20},
we assume that detected people do not spread the disease.)
The spreading of infections is assumed to be governed by the
inhomogeneous, linear growth equation
\be
\left(\frac{\mathrm{d}U}{\mathrm{d}t}\right)(t)=
k(t)\,
U(t)
+
\Phi(t),
\label{eq: growthequation}
\ee
where $k(t)$ is the instantaneous growth rate and $\Phi(t)$ accounts
for infections arising from people crossing the national border. For
simplicity, we set this influx to zero in this paper, in which case
$k(t)=\dot{U}(t)/U(t)$ with the short-hand notation $\dot{U}(t)$
for the time derivative on the left-hand side of 
\eqref{eq: growthequation}.

An equation of the form (\ref{eq: growthequation})
is usually part of a more refined epidemiological model%
~\cite{ReviewSIRI,ReviewSIRII,ReviewSIRIII}
that accounts explicitly for the recovery or death of infected persons.
For our purpose, the effect of these has been lumped into an overall
time-dependence of the rate $k(t)$. For example,
it evolves as the number of immune people grows,
restrictive measures change, mobility is affected, new
tracking systems are implemented, hospitals reach their capacity,
testing is increased, etc. Nevertheless, over a short period of time
where such conditions remain constant, and the fraction of immune
people does not change significantly, we can assume the effective
growth rate $k(t)$ to be piecewise constant in time.%
~\footnote{
Replacing the function $k(t)$, assumed to be differentiable, by
a piecewise constant function is a good approximation provided
$k(t)/\dot{k}(t)\gg\Delta t(k)$
where $\dot{k}(t)$ is the time derivative of $k(t)$
and $\Delta t(k)$ is given by Eq.\ (\ref{Deltat1})
with the replacement $k^{\,}_{1}\to k(t)$.
          }
We will exploit this below.

\subsection{Modeling intervention strategies}
\label{ssMIS}

For $t<0$, we assume a situation that is under control,
with a negative growth rate
\begin{subequations}
\label{eq: Modeling intervention strategies}
\be
k(t<0)\equiv
k^{\,}_{0}<0,
\label{eq: Modeling intervention strategies a}
\ee
as is the case in Switzerland since the lockdown in March,
with $k^{\,}_{0}\approx -0.07\,\mathrm{day}^{-1}$,
according to the estimates of Ref.~\cite{FergusonReport13}.
Such a stable state needs to be reached before a reboot of the economy can be
considered. At $t=0$ restrictive measures are first relaxed,
resulting in an increase of the growth rate $k$ from
$k^{\,}_{0}$ to $k^{\,}_{1}$, which we assume positive,
\be
k(t=0)=k^{\,}_{1}>0.
\label{eq: Modeling intervention strategies b}
\ee
Hence, compensating counter measures are required at later times
to avoid another exponential growth of the pandemic.

We now want to monitor the performance of policy strategies that relax
or re-impose restrictions, step by step. The goal for an optimal
policy is to reach a marginally stable state
(\ref{eq: definition marginal stable state})
(i.e., with $k=0$) as smoothly, safely, and rapidly as possible.
In other words, marginal stability is to be reached with the least possible
damage to health, economy, and society. This expected outcome is to be
optimized while controlling the risk of rare fluctuations.

To model the performance of policy strategies,
we neglect the contributions to the time evolution of
$k(t)$ due to the increasing immunity or the evolution in the age
distribution of infected people. We also neglect periodic temporal
fluctuations of $k(t)$
(e.g., due to alternation between workdays and weekends),
which can be addressed in further elaborations.
Instead, we assume that $k(t)$ changes only in response to policy
measures which are taken at specific times when certain criteria are
met, as defined by a policy strategy. An intervention is made when
the sampled testing data indicates that with high likelihood, $k(t)$
exceeds some upper threshold
\be
\kappa^{\,}_{+}\geq 0.
\label{eq: Modeling intervention strategies c}
\ee
Likewise,
a different intervention is made should $k(t)$ be detected to fall below some
negative threshold
\be
\kappa^{\,}_{-}\leq0.
\label{eq: Modeling intervention strategies d} 
\ee
Note that if there is 
substantial infection influx $\Phi(t)$ across the national borders, one
may want to choose the threshold $\kappa^{\,}_{+}$ to be negative, to
avoid a too large response to the influx. From now on we neglect the
influx of infections, and consider a homogeneous growth equation.

To reach decisions on policy measures, data is acquired by
daily testing of random sets of people for infections. We
assume that the tests are carried out at a limited rate $r$
(a finite number of tests divided by a nonvanishing unit of time).
Let $i(t,{\Delta t})$ be the fraction of positive infections detected among
the $r\,{\Delta t}\gg1$ tests carried out in the time interval
$[t,t+{\Delta t}]$. By the law of large numbers,
it is a Gaussian random variable with mean 
\be
\langle i(t,{\Delta t})\rangle=
\frac{\overline{U(t)}}{N},
\qquad
\overline{U(t)}\equiv
\int\limits_{t}^{t+{\Delta t}}\frac{\mathrm{d}t'}{{\Delta t}}\,
U(t')
\label{eq: Modeling intervention strategies e} 
\ee
and standard deviation
\be
\langle[i(t,{\Delta t})]^{2}\rangle^{1/2}_{\mathrm{c}}=
\sqrt{\frac{\langle i(t)\rangle}{r\,{\Delta t}}}=
\sqrt{\frac{\overline{U(t)}}{N\,r\,{\Delta t}}}.
\label{eq: Modeling intervention strategies f}
\ee
The current value of $k(t)$ is estimated as $k^{\mathrm{fit}}_{}(t)$
by fitting these test data to an exponential,
\textit{where only data since the last policy change should be used}. 
The fitting also yields the statistical uncertainty (standard deviation),
which we call $\delta k(t)$.
It will take at least 2-3 days to make a fit that is reasonably trustworthy.

If the instability threshold is surpassed by a certain level, i.e., if
\be
k^{\mathrm{fit}}(t)-\kappa^{\,}_{+}>\alpha\,\delta k(t)
\label{eq: Modeling intervention strategies g}
\ee
a new restrictive intervention is taken. If instead
\be
\kappa^{\,}_{-}-k^{\mathrm{fit}}(t)>\alpha\,\delta k(t)
\label{eq: Modeling intervention strategies h}
\ee
a new relaxing intervention is taken. Here, the parameter $\alpha$ is
a key parameter defining the policy strategy.  It determines the
confidence level
\be
p\equiv[1+ \mathrm{erf}(\alpha)]/2
\label{eq: Modeling intervention strategies i}
\ee
that policymakers require, before deciding to declare that a stability
threshold has indeed been crossed.
This strategy will result in a series of intervention times
\be
0\equiv
t^{\,}_{1}<t^{\,}_{2}<t^{\,}_{3}\cdots
\label{eq: Modeling intervention strategies j}
\ee
starting with the initial step to reboot at $t^{\,}_{1}=0$.
In the time window $[t^{\,}_{\iota},t^{\,}_{\iota+1}]$,
the growth rate
$k(t)$
is constant and takes the value
\be
k^{(\iota)}= k^{(\iota-1)}-\Delta k^{(\iota)},
\qquad
\iota=1,2,\cdots
\label{eq: Modeling intervention strategies k}
\ee
where a policy choice {with}
$\Delta k^{(\iota)}>0$
(corresponding to a restrictive measure) is made
to bring back $k(t)$ below the upper threshold $\kappa^{\,}_{+}$,
while {a} policy choice {with}
$\Delta k^{(\iota)}<0$
is made to bring back $k(t)$ above the lower threshold $\kappa^{\,}_{-}$.

The difficulty for policymakers is due to the fact that so far the
quantitative effect of an intervention is not known.
We model this uncertainty by
assuming $\Delta k^{(\iota)}$ to be random to a certain degree.

If at time $t$, $k^{\mathrm{fit}}(t)$ crosses the upper threshold
$\kappa^{\,}_{+}$ with confidence level $p$, we set $t^{\,}_{\iota}=t$
and a restrictive measure is taken, i.e.,
$\Delta k^{(\iota)}$ is chosen positive.
We take the associated decrement
$\Delta k^{(\iota)}$ to be uniformly distributed on the interval
\be
\left[
b\,\Delta k^{(\iota)}_{\mathrm{opt},+},
\frac{1}{b}\Delta k^{(\iota)}_{\mathrm{opt},+}
\right],
\label{eq: Modeling intervention strategies l}
\ee
where the optimum choice $\Delta k^{(\iota)}_{\mathrm{opt},+}$ is defined by
\be
\Delta k^{(\iota)}_{\mathrm{opt},+}\equiv
k^{\mathrm{fit}}\,
\left(t^{\,}_{\iota}\right)
-
\kappa^{\,}_{+}>0.
\label{eq: Modeling intervention strategies m}
\ee
The parameter {$b< 1$} describes the uncertainty about
the effects of the measures taken by policymakers.
While the policymakers aim to reset the growth factor $k$
to $\kappa^{\,}_{+}$, the result of the measure taken may range from having an
effect that is too small by a factor of $b$ 
to overshooting by a factor of $1/b$.
A measure with effect
$\Delta k^{(\iota)}=\Delta k^{(\iota)}_{\mathrm{opt},+}$
would be  optimal according to the best current estimate.
The larger $1-b$, the larger the uncertainty.
Unless stated otherwise, we assume $b=0.5$.

If instead $k^{\mathrm{fit}}(t)$ crosses the lower threshold
$\kappa^{\,}_{-}$ with confidence level $p$ at time $t$, 
we set $t^{\,}_{\iota}=t$
and a relaxing measure is taken, i.e.,
$\Delta k^{(\iota)}$ is chosen negative.
Again, $\Delta k^{(\iota)}$ is uniformly distributed on the interval
\be
\left[
-\frac{1}{b}\,\Delta k^{(\iota)}_{\mathrm{opt},-},
-b\, \Delta k^{(\iota)}_{\mathrm{opt},-}
\right]
\label{eq: Modeling intervention strategies n}
\ee
with the optimum choice $\Delta k^{(\iota)}_{\mathrm{opt},-}$ defined by
\be
\Delta k^{(\iota)}_{\mathrm{opt},-}\equiv
\kappa^{\,}_{-}
-
k^{\mathrm{fit}}\,
\left(t^{\,}_{\iota}\right)>0.
\label{eq: Modeling intervention strategies o}
\ee
\end{subequations}

The  process described above is stochastic for two reasons. First, the
sampling comes with the usual uncertainties in the law of large
numbers. Second, the effect of policy measures is not known beforehand
(even though it may be learnt in the course of time, which we do not
include here). It should be clear that the faster the testing the more
rapidly one can respond to a super-critical situation.

A significant simplification of the model occurs when the two
thresholds are chosen to vanish,
\begin{subequations}
\label{eq: kappas are zero}
\be
\kappa^{\,}_{\pm}=0,
\label{eq: kappas are zero a}
\ee
in which case
\be
k^{(\iota)}= k^{(\iota-1)}-\Delta k^{(\iota)},
\qquad
\iota=1,2,\cdots,
\label{eq: kappas are zero b}
\ee
with
$|\Delta k^{(\iota)}|$ uniformly distributed on the interval
\be
\left[
b\, |k^{\mathrm{fit}}(t^{\,}_{\iota})|,
\frac{1}{b}\,|k^{\mathrm{fit}}(t^{\,}_{\iota})|
\right].
\label{eq: kappas are zero c}   
\ee
\end{subequations}
In this case the system will usually tend to a critical steady state with
$k(t\to\infty)\to 0$, as we will show explicitly below.  In this case
the policy strategy can simply be rephrased as follows. As soon as one
has sufficient confidence that $k$ has a definite sign, one
intervenes, trying to bring $k$ back to zero. The only parameter
defining the strategy is $\alpha$.

\subsection{Testing and fitting procedure}

Let us now detail the fitting procedure and analyze the typical time
scales involved between subsequent policy interventions
when choosing the thresholds (\ref{eq: kappas are zero}).
After a policy change at time $t^{\,}_{\iota}$, data is acquired over a time
window ${\Delta t}$.  We then proceed with the following steps to
estimate the time $t^{\,}_{\iota+1}$ at which the next policy change
must be implemented.

\textit{Step 1: Measurement}
We split the time window
\begin{subequations}
\label{eq: step 1}
\be
\Delta T^{\,}_{\iota}\equiv
[t^{\,}_{\iota},t^{\,}_{\iota}+{\Delta t}]
\label{eq: step 1 a}
\ee
of length ${\Delta t}$ after the policy change
into the time interval
\be
\Delta T^{\,}_{\iota,1}\equiv
\left[t^{\,}_{\iota},t^{\,}_{\iota}+\frac{{\Delta t}}{2}\right]
\label{eq: step 1 b}
\ee
and the time interval
\be
\Delta T^{\,}_{\iota,2}\equiv
\left[t^{\,}_{\iota}+\frac{{\Delta t}}{2},t^{\,}_{\iota}+{\Delta t}\right].
\label{eq: step 1 c}
\ee
Testing delivers the number of currently infected people
\be
N^{\,}_{\iota,1}({\Delta t})=
r\,{\Delta t}\,i(t^{\,}_{\iota},\frac{{\Delta t}}{2})
\ee
for the time interval (\ref{eq: step 1 b}) 
and
\be
N^{\,}_{\iota,2}({\Delta t})=
r\,{\Delta t}\,i(t^{\,}_{\iota}+\frac{{\Delta t}}{2},\frac{{\Delta t}}{2})
\ee
for the time interval (\ref{eq: step 1 c}), where we recall that $r$ denotes
the number of people tested per unit time. Given those two measurements over
the time window ${\Delta t}/2$,
we obtain the estimate
\be
k^{\mathrm{fit}}_{\iota}({\Delta t})=
\frac{2}{{\Delta t}}\,
\ln\left(\frac{N^{\,}_{\iota,2}({\Delta t})}{N^{\,}_{\iota,1}({\Delta t})}\right)
\label{eq: step 1 d}
\ee
with the standard deviation 
\be
\delta k({\Delta t})=
\frac{2}{{\Delta t}}\,
\sqrt{
\frac{1}{N^{\,}_{\iota,1}({\Delta t})}
+
\frac{1}{N^{\,}_{\iota,2}({\Delta t})}
     },
\label{eq: step 1 e}
\ee
\end{subequations}
as follows from the statistical uncertainty
$\sqrt{N^{\,}_{\iota,\gamma}(\Delta t)}$ of the sampled numbers
$N^{\,}_{\iota,\gamma}(\Delta t)$ and standard error propagation.
The above recipe can be replaced by a more sophisticated
Levenberg-Marquardt fitting procedure, which yields more accurate estimates
for $k(t)$ with a smaller uncertainty $\delta k(t)$.
We have confirmed that this uniformly improves the performance of the
mitigation strategy.

\textit{Step 2: Condition for new policy intervention}
A new policy intervention is taken once the magnitude
$|k^{\mathrm{fit}}_{\iota}({\Delta t})|$
with $k^{\mathrm{fit}}_{\iota}({\Delta t})$
given by Eq.\ (\ref{eq: step 1 d}) 
exceeds
$\alpha\,\delta k({\Delta t})$
with $\delta k({\Delta t})$
given by Eq.\ (\ref{eq: step 1 e}).
Here, $\alpha$ controls the accuracy to which the actual
$k$ has been estimated at the time of the next intervention.
The condition 
\begin{subequations}
\label{Step 3: CNPI}
\be
|k^{\mathrm{fit}}_{\iota}({\Delta t})|=
\alpha\,\delta k({\Delta t}),
\label{Step 3: CNPI a}
\ee
for a new policy intervention thus becomes
\be
\left|
\ln\left(\frac{N^{\,}_{\iota,2}({\Delta t})}{N^{\,}_{\iota,1}({\Delta t})}\right)
\right|=
\alpha\,
\sqrt{
\frac{1}{N^{\,}_{\iota,1}({\Delta t})}
+
\frac{1}{N^{\,}_{\iota,2}({\Delta t})}
     }.
\label{Step 3: CNPI b}
\ee
\end{subequations}

\textit{Step 3: Comparison with modeling}
We call  ${i(t)=U(t)/N}$ the actual fraction of infections (in the entire
population) as a function of time, which we assume to follow a simple
exponential evolution between two successive policy interventions, i.e.,
the normalized solution 
\be
i(t^{\,}_{\iota}+t')=i(t^{\,}_{\iota})\,\exp(k^{\,}_{\iota}\,t')
\ee
to the growth equation (\ref{eq: growthequation}) on the interval
$t^{\,}_{\iota}<t'<t^{\,}_{\iota+1}$.
The expected number of newly detected infected people in the time interval
(\ref{eq: step 1 b}) 
is
\begin{subequations}
\label{eq: Step 2: Comparison with modeling}
\begin{align}
\langle
N^{\,}_{\iota,1}({\Delta t})
\rangle=&\,
r\,
\int\limits_{0}^{{\Delta t}/2}\mathrm{d}t'\,
i(t^{\,}_{\iota}+t')
\nonumber\\
=&\,
r\,
i(t^{\,}_{\iota})\,
\frac{
e^{k^{\,}_{\iota}\,{\Delta t}/2}-1
     }
     {
k^{\,}_{\iota}
     }.
\label{eq: Step 2: Comparison with modeling a}
\end{align}
Similarly,
the predicted number of infected people in the time interval
(\ref{eq: step 1 c}) 
is
\begin{align}
\langle
N^{\,}_{\iota,2}({\Delta t})
\rangle=&\,
r\,
\int\limits_{{\Delta t}/2}^{{\Delta t}}\mathrm{d}t'\,
i(t^{\,}_{\iota}+t')
\nonumber\\
=&\,
r\,
i(t^{\,}_{\iota})\,
\frac{
e^{k^{\,}_{\iota}\,{\Delta t}/2}
\left(  
e^{k^{\,}_{\iota}\,{\Delta t}/2}-1
\right)
     }
     {
k^{\,}_{\iota}
     }.
\label{eq: Step 2: Comparison with modeling b}
\end{align}
\end{subequations}

\textit{Step 4: Estimated time for a new policy intervention}
We now approximate
$N^{\,}_{\iota,1}$
and
$N^{\,}_{\iota,2}$
by replacing them with their expectation value
Eqs.\
(\ref{eq: Step 2: Comparison with modeling a})
and
(\ref{eq: Step 2: Comparison with modeling b}),
respectively, and anticipating the limit
\begin{subequations}
\label{Step 4}
\be
k^{\,}_{\iota}\,{\Delta t}/2\ll1.
\label{Step 4: a}
\ee
We further anticipate that for safe strategies the fraction of
currently infected people $i(t)$ does not vary strongly over time.
More precisely, it hovers around the value $i^{*}$ defined in
Eqs.\ (\ref{eq: def ic}) and (\ref{eq: estimate i*})
(see Fig.\ \ref{fig: comparison}).
We thus insert
\be
N^{\,}_{\iota,1}\approx
N^{\,}_{\iota,2} \approx
r\,i(t^{\,}_{\iota})\,{\Delta t}/2\approx 
r\, i^{*}\,{\Delta t}/2
\label{Step 4: b}
\ee
into Eq.\ (\ref{Step 3: CNPI b}) and solve for ${\Delta t}$. 
The solution is the time until the next intervention
\be
\Delta t^{\,}_{\iota}\equiv
t^{\,}_{\iota+1}-t^{\,}_{\iota}=
\frac{(4\,\alpha)^{2/3}}{(k^{2}_{\iota}\,r\,i^{*})^{1/3}},
\label{Step 4: c}
\ee
from which we deduce the relative increase 
\be
\begin{split}
\frac{i(t^{\,}_{\iota+1})}{i(t^{\,}_{\iota})}\equiv&\,
\exp\left(k^{\,}_{\iota}\,\Delta t^{\,}_{\iota}\right)
\\
=&\,
\exp
\left(
\mathrm{sgn}(k^{\,}_{\iota})\,
(4\,\alpha)^{2/3}\,\left(\frac{|k^{\,}_{\iota}|}{r\,i^{*}}\right)^{1/3}
\right)
\label{Step 4: d}
\end{split}
\ee
\end{subequations}
of the fraction of currently infected people over the time window.
This relative increase is close to $1$ if the argument of
the exponential on the right-hand side is small.

We will show below that the characteristics
\begin{subequations}
\be
\Delta t^{\,}_{1}=
\frac{(4\,\alpha)^{2/3}}{(k^{2}_{1}\,r\,i^{*})^{1/3}},
\label{Deltat1}
\ee
and 
\be
\label{initialincrease}
\frac{i(t^{\,}_{2})}{i(t^{\,}_{1})}=
\exp
\left(
(4\,\alpha)^{2/3}\,\left(\frac{k^{\,}_{1}}{r\,i^{*}}\right)^{1/3}
\right)
\ee
\end{subequations}
of the first time interval  $[t^{\,}_{1},t^{\,}_{2}]$
set the relevant scales for the entire process.
From Eqs.\ (\ref{Step 4: c}) and (\ref{Step 4: d}),
we  infer the following important result.
The higher the testing frequency $r$,
the smaller the typical variations in the
fraction of currently infected people, and thus in the case numbers.
The band width of fluctuations decreases as $r^{-1/3}$
with the testing rate.

\subsubsection{Critical fraction of infections}

As one should expect, it is always the average rate to
detect a currently infected person, $r\,i^{*}$, which enters
into the expressions
(\ref{Step 4: c}) and (\ref{Step 4: d}). The higher the fraction $i^{*}$,
the more reliable is the sampling, the shorter is the time to converge
toward the marginal state (\ref{eq: definition marginal stable state}),
and the smaller are the fluctuations of the
fraction of infected people.

If the fraction $i^{*}$ is
too low the statistical fluctuations become too large and little
statistically meaningful information can be obtained.
On the other hand, if the fraction of infections drops to
much lower values, then policy can be considered to have
been successful and can be maintained until further tests show otherwise.

We seek an upper bound for a manageable $i^{*}$.
Here we consider the parameters of Switzerland.
However, they can easily be adapted to any other country.
We assume that a
fraction $p^{\mathrm{CH}}_{\mathrm{ICU}}$ of infected people in Switzerland
needs to be in intensive care.~\footnote{More precisely,
$p^{\mathrm{CH}}_{\mathrm{ICU}}$ is the expected
time (in Switzerland) for an infected person to spend
in an intensive care unit (ICU)
divided by the expected time to be sick.} Here, we
will use the value $p^{\mathrm{CH}}_{\mathrm{ICU}}=0.05$. 
Let $\rho^{\,}_{\mathrm{ICU}}$ be the number of ICU beds
per inhabitant that shall be allocated to COVID-19 patients.
The Swiss national average is about%
~\cite{ICU beds in CH}
\begin{subequations}
\be
\rho^{\mathrm{CH}}_{\mathrm{ICU}}\approx
\frac{1200}{8'500'000}\approx
1.4\times10^{-4}.
\ee
For the pandemics not to overwhelm the health system, one thus needs
to maintain the fraction of currently infected people safely below
\be
i(t)\leq i^{\,}_{\mathrm{c}}=
\frac{\rho^{\mathrm{CH}}_{\mathrm{ICU}}}{p^{\mathrm{CH}}_{\mathrm{ICU}}}=
0.0028,
\label{eq: def ic}
\ee
\end{subequations}
together with similar constraints related to the
capacity for hospitalizations, medical care personnel and equipment
for specialized treatments.
We take the  constraint from intensive care units
to obtain an order of magnitude for the upper limit admissible
for the infected fraction of people, $i$. 
A recent study based on random testing reports that
the fraction of people currently infected with the virus
lies within the confidence interval $[0.0012,0.0076]$
in Austria (whereby half of the infected people in the sample
were previously undetected) \cite{Austria}.
The estimates in Ref.~\cite{FergusonReport13} suggest
that the fraction of acutely infected people was even close to $0.01$
before the lock-down in Switzerland.
This indicates that our threshold estimate
(\ref{eq: def ic}) is conservative.
If the actual threshold (which depends on the country,
the structure of its population, and its health-care infrastructure) is
higher, the testing frequency required to reach a defined accuracy
decreases in proportion. 

The objective is to mitigate the pandemic so that values of the order
of $i^{\,}_{\mathrm{c}}$ or below are achieved. Before that level is reached
restrictions cannot be relaxed.  It may prove difficult to push the
fraction of infected people significantly below $i^{\,}_c$, since the
recent experience in most European countries suggests that it is very
hard to ensure that growth rates $k$ fall well below 0. The main aim
would then be to reach at least stabilization of the number of
currently infected people ($k=0$).

For the following we thus assume that the fraction of infections $i$
will stagnate around a value $i^{*}$ of the order of
$i^{\,}_{\mathrm{c}}$. We will discuss below what ratio
$i^{*}/i^{\,}_{\mathrm{c}}$ can be considered safe.

\subsection{Required testing rate}

We seek the testing rate that is needed to obtain a strategy
with satisfactory outcome. We assume that after the reboot at $t^{\,}_{1}=0$,
the initial growth rate may turn out to be fairly high, say of the order
of the unmitigated growth rate.  In many
European countries a doubling of cases was observed every three days
before restrictive measures were introduced. This corresponds to a
growth rate of
\begin{subequations}
\be
k^{\,}_{0}=
\frac{\ln(2)}{3\,\mathrm{days}}\approx
0.23\,\mathrm{day}^{-1}.
\ee
We assume an initial  growth rate of
\be
k^{\,}_{1}=0.1\,{\mathrm{day}^{-1}}
\ee
just after the reboot.
We choose the reasonably stable confidence parameter
\be
\alpha=3.
\ee
In Sec.\ \ref{sec: Simulation of mitigation ...}
we will find that this choice strikes a
good balance between several performance criteria
(see Fig.~\ref{fig: Performance fct alpha}).
We further assume that the rate of infections initially stagnates at a level
of (for Switzerland)
\be
i^{*}=
\frac{i^{\,}_{\mathrm{c}}}{4}\approx 0.0007.
\label{eq: estimate i*}
\ee 
The level $i^{*}$ should, however, be
measured by random testing before a reboot is attempted. 
We should then ensure that the first relative increase of
\be
\frac{i(t^{\,}_{2})}{i(t^{\,}_{1})}=\frac{i(t^{\,}_{2})}{i(0)}
\ee
\end{subequations}
does not exceed a factor of 4.
From Eq.~(\ref{initialincrease}), we thus obtain the requirement 
\be
r\geq r^{\,}_{\mathrm{min}} \equiv 
\frac{(4\,\alpha)^2}{(\ln 4)^{3}}\,
\frac{k^{\,}_{1}}{i^{*}}\approx
7'700\,\mathrm{day}^{-1}
\label{testfrequency}
\ee
for the testing rate $r$. Note the inverse proportionality
to the parameter $i^{*}$,
for which Eq.\ (\ref{eq: estimate i*}) is a conservative estimate.
Using this value yields an estimate of the order of
magnitude required for Switzerland.
In the next section we simulate a full mitigation
strategy and confirm that with additional capacity for just about
15'000 random infection tests per day a nation-wide, safe reboot can
be envisioned for Switzerland.

We close with two observations.
First, this minimal testing frequency is just twice the
testing frequency presently available for suspected infections
and medical staff in Switzerland.
Second, while the latter tests require a high sensitivity with as few
false negatives as possible, random testing can very well be carried
out with tests of lower quality in that respect. Indeed, an increase
in false negatives acts as a systematic error in the estimate of the
infected fraction of people, which, however, drops out in the
determination of its growth rate,%
~\footnote{
If the infected fraction of people is $i(t)$, its growth rate is
$\dot{i}(t)/i(t)\equiv k(t)$ with the time derivative of $i(t)$
denoted by $\dot{i}(t)$.
          }
as long as the fraction $i$ is not close to 1.
However, the success of random testing does rely on a very
low probability ($\ll i^{*}$) of false positives
(as is the case of current PCR tests).
Otherwise the signal from true positives would rapidly be
overwhelmed by the noise from false positives.

\subsection{Further intervention steps after the reboot}

After the reboot at time $t^{\,}_{1}=0$ further interventions will be
necessary, as we assume that the reboot will have resulted in a
positive growth rate $k^{\,}_{1}$. In subsequent interventions, the
policymakers try to take measures that aim at reducing the growth rate
to zero. Even if they had perfect knowledge of the current
growth rate $k(t)$,
they would not succeed immediately since they do not know
the precise quantitative effect of the measures they will take.
Nevertheless, had they complete knowledge of $k(t)$,
our model assumes that they would be able to gauge their intervention
such that the actual effect on $k(t)$ differs at most by a factor between
$b$ and $1/b$ from the targeted value,
which would reduce $k(t)$ to 0.
This and the assumption $b\geq 0.5$ implies that, if $\alpha$ is large,
so that $k(t)$ is known with relatively high
precision at the time of intervention,
the growth rate $k^{\,}_{2}$ is smaller than $k^{\,}_{1}$ in magnitude
with high probability (tending rapidly to $1$ as $\alpha\to\infty$).%
~\footnote{
One uses Eq.\ (\ref{eq: kappas are zero})
to reach this conclusion.
          }
The smaller $\alpha$ however,
the more likely it becomes, that $k(t)$ is overestimated, and an exaggerated corrective measure is taken, which may destabilize the system.
In this context, we observe that the ratio 
\be
0<\rho^{\,}_{\iota}\equiv\frac{|k^{\,}_{\iota}|}{|k^{\,}_{\iota-1}|}<\infty
\ee
is a random variable with a distribution that is independent of
$\iota$ in our model.
To proceed, we assume that $\alpha$ is sufficiently large, 
such that the probability that $\rho^{\,}_{\iota}<1$ is indeed high.

The second policy intervention occurs after a time
that can be predicted along the same lines that
lead to Eq.~(\ref{Step 4: c}). One finds
\be
\Delta t^{\,}_{2}\approx
\Delta t^{\,}_{1}\,
\left(
\frac{|k^{\,}_{2}|}{|k^{\,}_{1}|}
\right)^{-2/3},
\ee
where $\Delta t^{\,}_{1}$ is given by Eq.\
(\ref{Deltat1}).
Since, the growth rate $k^{\,}_{3}$ is likely to be smaller than
$k^{\,}_{2}$ in magnitude, the third intervention takes place
after yet a longer time span, etc.  If we neglect that the fitted
value $k^{\mathrm{fit}}_{\iota}(t)$ differs slightly from $k^{\,}_{\iota}$
({a difference that}
is negligible when $\alpha\gg1$), our model ensures that
$k^{\,}_{\iota}/k^{\,}_{\iota-1}$ is uniformly distributed in $[b-1,1/b-1]$.
After the $\iota$-th intervention the growth rate is down in magnitude to
\be
|k^{\,}_{\iota}|=
|k^{\,}_{0}|\,
\prod_{\iota'=1}^{\iota}
\rho^{\,}_{\iota'}.
\ee
To reach a low final growth rate $k^{\,}_{\mathrm{final}}$,
a typical number 
$n^{\,}_{\mathrm{int}}(k^{\,}_{\mathrm{final}})$
of interventions are required after the reboot,
where 
\be
\label{interventions}
n^{\,}_{\mathrm{int}}(k^{\,}_{\mathrm{final}})\approx
\frac{
\ln \frac{|k^{\,}_{\mathrm{final}}|}{|k^{\,}_{1}|}
     }
     {
\langle\ln\rho^{\,}_{\iota}\rangle
     }=
C(b)\,
\ln\frac{|k^{\,}_{1}|}{|k^{\,}_{\mathrm{final}}|},
\ee
where the constant
$C(b)=-1/\langle\ln\rho^{\,}_{\iota}\rangle$
depends on the policy uncertainty parameter $b$.

The time to reach this low rate is dominated by
the last  time interval {which yields} the estimate
\be
\label{timescale}
T(k^{\,}_{\mathrm{final}})\sim
\Delta t^{\,}_{n^{\,}_{\mathrm{int}}(k^{\,}_{\mathrm{final}})}\approx
\left(
\frac{|k^{\,}_{1}|}{|k^{\,}_{\mathrm{final}}|}
\right)^{2/3}\,
\Delta t^{\,}_{1}\,.
\ee
Thus, the system converges to the critical state where $k=0$, but
never quite reaches it. At late times $T$, the residual growth rate
behaves as $k^{\,}_{\mathrm{final}}\sim T^{-3/2}$.

\subsection{Choosing an optimal intervention strategy}

The parameter $\alpha$ encodes
the confidence policymakers need 
about the present state before they take a decision. Here we discuss
various measures that allow choosing an optimal value for $\alpha$.

As $\alpha$ decreases starting from large values, the time for
interventions decreases, being proportional to $\alpha^{2/3}$
according to Eq.\ (\ref{Deltat1}).
Likewise the fluctuations of infection numbers
will initially decrease. However, the logarithmic average
$-\langle\ln\rho^{\,}_{\iota}\rangle$ in the denominator of
Eq.~(\ref{interventions}) will also decrease, and thus the
necessary number of interventions increases. Moreover, when $\alpha$
falls below $1$, interventions become more and more
ill-informed and erratic. It is not even obvious anymore
that the marginally stable
state is still approached asymptotically. From these two limiting
considerations, we expect 
\be
\alpha=O(1)
\label{eq: optimum choice for alpha}
\ee
to be an optimal choice for $\alpha$.

Let us now discuss a few quantitative measures of the performance of
various strategies, which will allow policymakers to make an optimal
choice of confidence parameter for the definition of a mitigation strategy.

\subsubsection{Time scale  to approach the marginal state} 

The time to reach a certain level of quiescence (low growth rates,
infrequent interventions) is given by the time (\ref{timescale}), and
thus by the expectation value of $\Delta t^{\,}_{1}$.

\subsubsection{Political cost}

As a measure for the political cost, $C^{\,}_{\mathrm{P}}$,
we may {consider the number of interventions that have to be taken}
to reach quiescence.  As we saw in Eq.~(\ref{interventions}),
it scales inversely with the logarithmic average
of the ratios of growth rates, $\rho$, i.e.,
\be
\label{CP}
C^{\,}_{\mathrm{P}}\propto
\left(\langle-\ln\rho^{\,}_{\iota}\rangle\right)^{-1}.
\ee

\subsubsection{Health cost}

If restrictions are over-relaxed, the infection numbers will grow with
time. The maximal fraction of currently infected people must never be allowed to
rise above the manageable threshold of $i^{\,}_{\mathrm{c}}$. This means that
continuous (random) monitoring of the fraction of infected people is
needed, so that given the knowledge from the time before the reboot,
about the conditions under which the system can be stabilized,
lockdown conditions can always be imposed at a time that is
sufficient to prevent reaching the level of $i^{\,}_{\mathrm{c}}$. Beyond this
consideration one may want to keep the expected maximal increase of
infection numbers low, which we take as a measure of health costs
$C^{\,}_{\mathrm{H}}$,
\be C^{\,}_{\mathrm{H}}\equiv
\label{CH}
\max_{t}\left\{\frac{i(t)}{i(0)}\right\}.
\ee
Note that as defined,
$C^{\,}_{\mathrm{H}}$ is a stochastic number.  Its mean and tail
distribution (for large $R$) will be of particular importance.

\subsubsection{Economic and social cost}

Imposing restrictions such that $k<0$ imply restrictions beyond what
is absolutely necessary to maintain stability. If we assume that the
economic cost $C^{\,}_{\mathrm{E}}$ is proportional to the excess negative growth
rate, $-k$ (and a potential gain proportional to $k$), a measure for
the economic cost is the summation over time of $-k(t)$,
\be
\label{CE}
C^{\,}_{\mathrm{E}}\propto
-\int\limits_{0}^{\infty}\mathrm{d}t\,k(t),
\ee
which converges, since
$k(t)$ decays as a sufficiently fast power law. Hereto, $C^{\,}_{\mathrm{E}}$ is a
stochastic variable that depends on the testing history and the policy
measures taken.  However, its mean and standard deviation could be used as indicators 
of economic performance.

\section{Simulation of mitigation strategy by random testing}
\label{sec: Simulation of mitigation ...} 

We introduced in Sec.\ \ref{sec: National modeling and intervention}
a feedback and control strategy to
tune to a marginal state with vanishing growth rate $k=0$ after an initial reboot.
Interventions were only taken based on the measurement of the
growth rate. However, in practice, a more refined strategy will be
needed. In case the infection rate drops significantly below $i^{*}$,
one might (depending on netting out political and economic pressures, something which the authors of this paper are not doing here) benefit from a positive growth rate $k$. We thus
assume that if $i(t)/i^{*}$ falls below some threshold $i^{\,}_{\mathrm{low}}=0.2$,
we intervene by relaxing some measures, that we assume to increase $k$ by
an amount uniformly distributed in $[0,k^{\,}_{1}]$, but without letting $k$
exceed the maximal value of $k^{\,}_{\mathrm{high}}=0.23$.  Likewise, one
should intervene when the fraction $i(t)$ grows too large. We do so
when $i(t)/i^{*}$ exceeds $i^{\,}_{\mathrm{high}}=3$. In such a situation we
impose restrictions resulting in a decrease of $k$ by a quantity
uniformly drawn from $[k^{\,}_{\mathrm{high}}/2,k^{\,}_{\mathrm{high}}]$. The precise
algorithm is given in the Supplementary Information.

Figure \ref{fig: algorithmic policy interv}
shows how our algorithm implements policy releases
and restrictions in response to test data. The initial infected
fraction and growth rate are $i(0)=i^{\,}_{\mathrm{c}}/4=0.0007$
and $k^{\,}_{1}=0.1$,
respectively, with a sampling interval of one day.
We choose $\alpha=3$ and a number of $r=15'000$ tests per day.
Figure \ref{fig: algorithmic policy interv}(a)
displays the infection fraction, $U(t)/N$,
as a function of time, derived using our simple exponential growth
model, which is characterized by a single growth rate that changes
stochastically at interventions [Eq.~(\ref{eq: growthequation}) without
the source term]. In the absence of intervention, the infected
population would grow rapidly representing uncontrolled runaway of a
second epidemic.
At each time step (day) the currently infected fraction of the
population is sampled.  The result is normally distribution with mean
and standard deviation given by Eqs.\
(\ref{eq: Modeling intervention strategies e}) and
(\ref{eq: Modeling intervention strategies f})
to obtain $i(t)$. The former are represented by small circles, the
latter by vertical error bars in
Fig.~\ref{fig: algorithmic policy interv}.
If $i/i^{*}$ lies outside the range
$[i^{\,}_{\mathrm{low}},i^{\,}_{\mathrm{high}}]$, we intervene as
described above. Otherwise, on each day $k^{\mathrm{fit}}(t)$ and
its standard deviation are estimated using the data since the last
intervention. With this, at each time step, Eqs.\
(\ref{eq: Modeling intervention strategies m})
to
(\ref{eq: Modeling intervention strategies o})
decide whether or not to intervene. In
Fig.~\ref{fig: algorithmic policy interv},
each red circle represents an intervention and
therefore either a decrease or increase of the growth rate constant
of our model.

\begin{figure}[t]
\begin{center}
\includegraphics[width=0.95\linewidth]{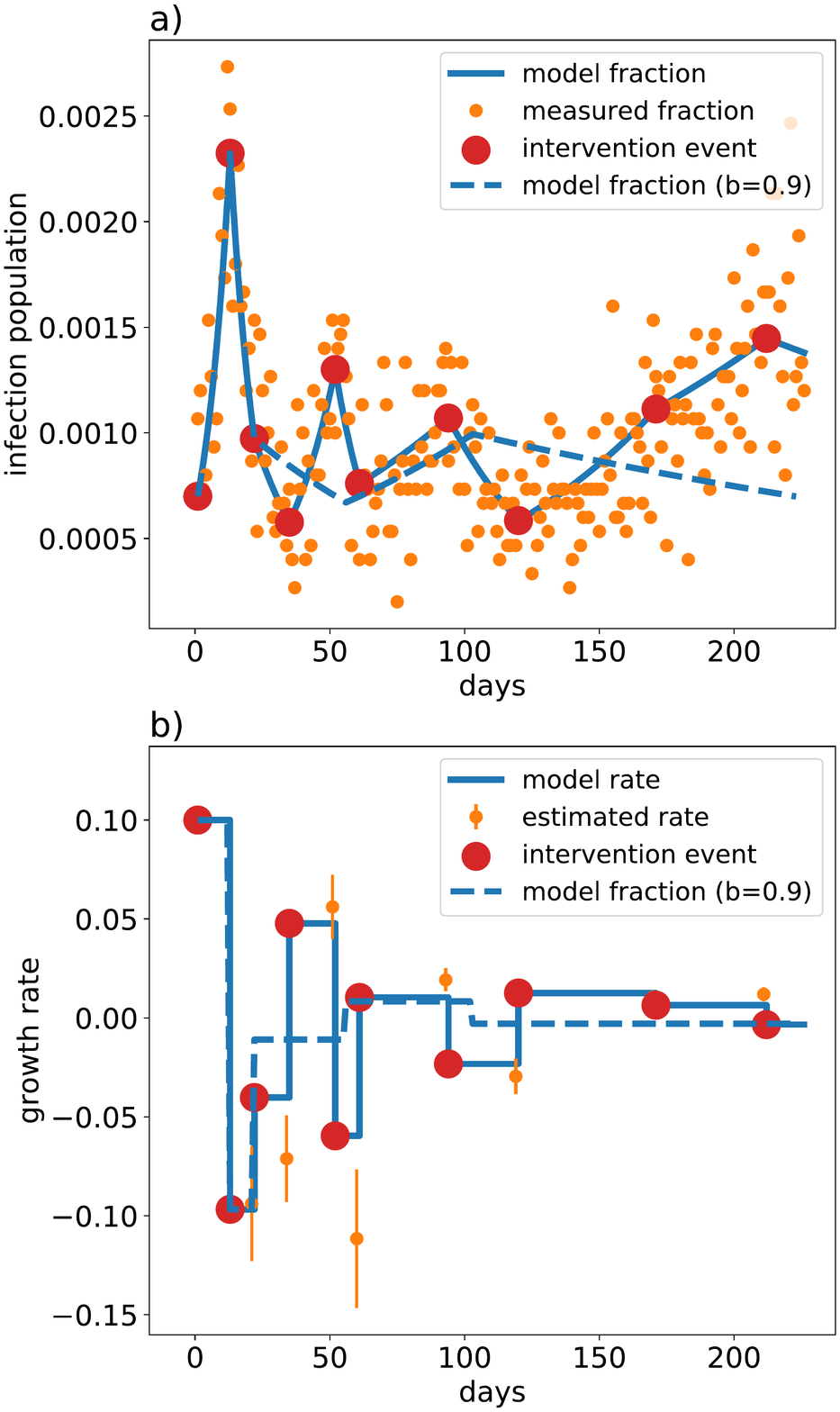}
\end{center}
\caption{
Our algorithm implements policy releases and restrictions aiming at
maintaining a vanishing growth rate. It intervenes whenever the
estimated slope of the fraction of infected people is found to be
non-zero, here with confidence level $\alpha=3$.
We plot the model infection
fraction $U(t)/N$ and the detected infection fraction $i(t)$ as a
function of days in panel (a). The model growth rate $k(t)$ (solid line) and
the estimated growth rate $k^{\,}_{\mathrm{est}}$ at times of intervention
are shown in panel (b) for the
parameters $i(0)=0.0012$, $k^{\,}_{1}=0.1$, and a test rate of
$r=15'000\,\mathrm{day^{-1}}$. The dashed blue line corresponds
to a history of interventions where we assumed that the effect of
policy interventions is better known
(described by an uncertainty parameter $b=0.9$, instead of $b=0.5$),
so that convergence is much faster.
       }
\label{fig: algorithmic policy interv}
\end{figure}

Figure \ref{fig: algorithmic policy interv}
shows the evolution of the fraction of currently
infected people. After an initial growth with rate $k^{\,}_{1}$
subsequent interventions reduce the growth rate down to low levels
within a few weeks.
At the same time the  fraction of infected people stabilizes 
at a scale similar to $i^{*}$. For the given parameter-set
this is a general trend independent of realization. 
Figure \ref{fig: algorithmic policy interv}(b)
displays the instantaneous value of the
model rate constant and also the estimated value together with its
fitting uncertainty. The estimate follows the model value reasonably
well. One sees that the interventions occur when the
uncertainty in $k$ is sufficiently small.

\begin{figure}[t]
\begin{center}
\includegraphics[width=0.95\linewidth]{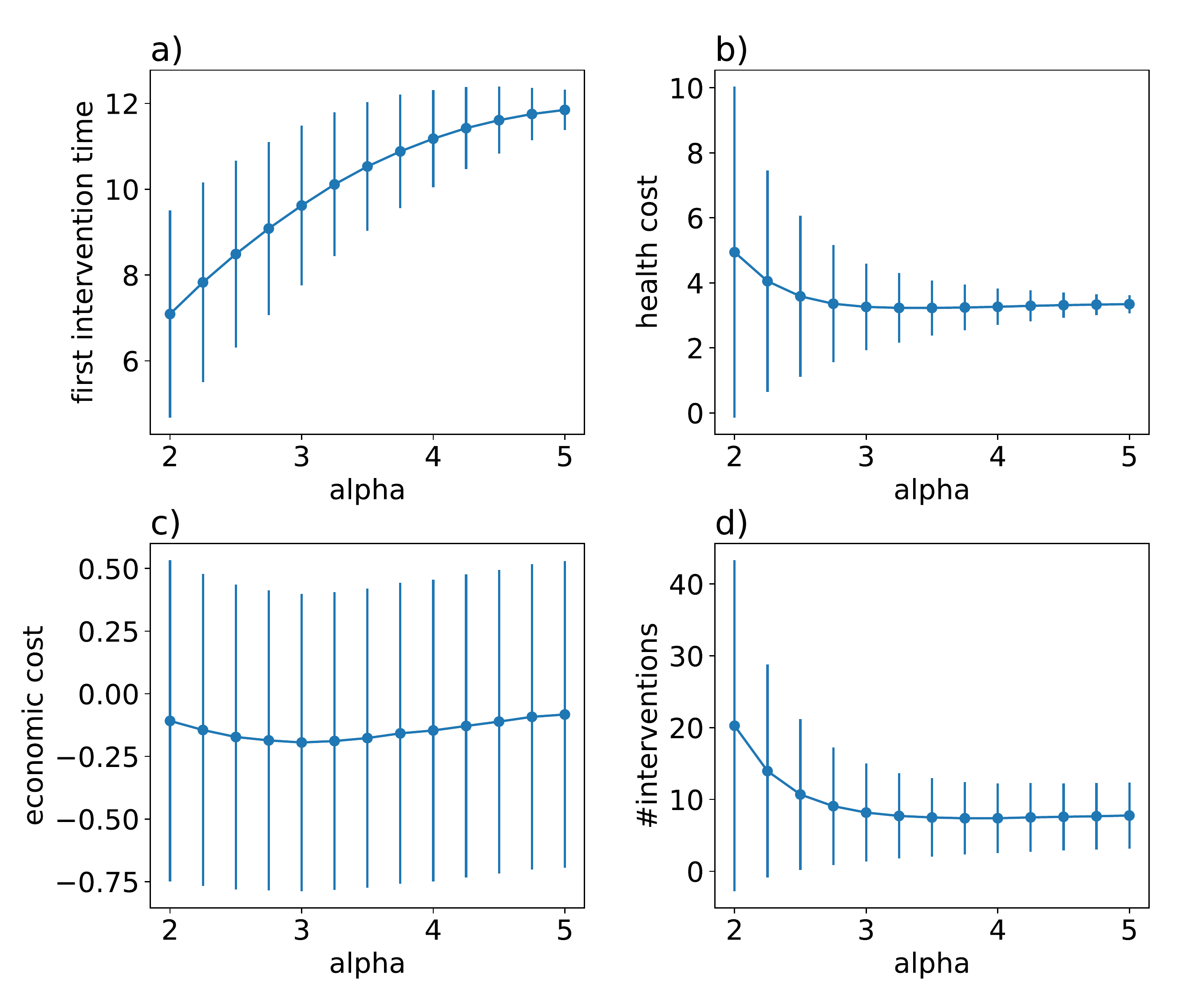}
\end{center}
\caption{
Performance of the mitigation strategy as a function
of the confidence parameter $\alpha$, for a number
$r=15'000$ tests per day  and an initial growth rate $k^{\,}_{1}=0.1$.
We plot the time scale $\Delta t^{\,}_{1}$ (a),
and the health (b), economic (c)  and political
(as measured by numbers of interventions to achieve a steady state) (d)
costs [Eqs.~(\ref{CP})--(\ref{CE})] as measures of performance.
The circles are the mean values,
the vertical lines indicate the standard deviations
of the respective quantities.
        }
\label{fig: Performance fct alpha}
\end{figure}

\subsection{Simulation results}

We now assume that we have the capacity for $r=15'000$ per day, and
assess the performance of our strategy as a function of the confidence
parameter $\alpha$ in Fig.\ \ref{fig: Performance fct alpha}. Values of
$\alpha\leq 2$ lead to rapid, but at the same time erratic
interventions, as is reflected by a rapidly growing number of
interventions.  For larger values of $\alpha$, the time scale to
reach a steady state increases while the economic and health costs
remain more or less stable. A reasonable compromise between minimizing
the number of interventions, and shortening the time to reach a steady
state suggests a choice of $\alpha\approx 2.5-3.5$.

\begin{figure}[h]
\begin{center}
\includegraphics[width=0.95\linewidth]{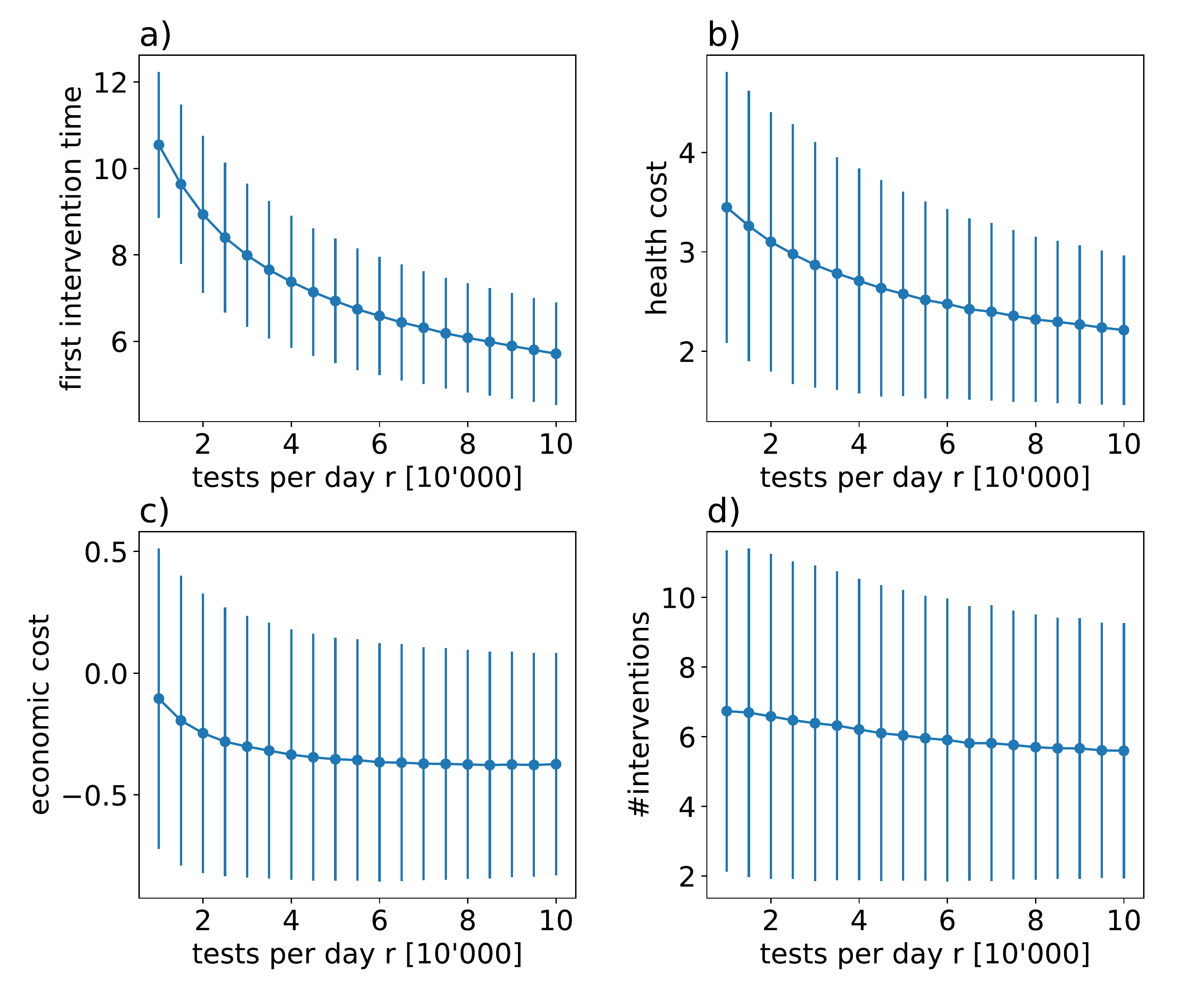}
\end{center}
\caption{Performance of the mitigation strategy as a function
of the number of tests $r$ per day, for a fixed value of $\alpha=3$
and an initial growth rate $k^{\,}_{1}=0.1$.
We plot the time scale $\Delta t^{\,}_{1}$ (a),
and the health (b), economic (c)  and political (as measured by numbers of interventions to achieve  a steady state (d) costs
[Eqs.~(\ref{CP})--(\ref{CE})] as measures of performance.
The circles are the mean values,
the vertical lines indicate the standard deviations
of the respective quantities.
The large uncertainties in the economic costs, e.g.,
are a consequence of  the relatively large uncertainty
in the effect of interventions ($b=0.5$). If the latter is better known,
the standard deviation of the cost functions will decrease accordingly.
        }
\label{fig: Performance fct r}
\end{figure}

It is intuitive that the higher the number $r$ of tests per day is,
the better the mitigation strategy will perform. The characteristic
time to reach a final steady state decreases as $r^{-1/3}$,
see Eq.~(\ref{Deltat1}). Other measures for performance 
improve monotonically upon increasing $r$. This is confirmed and quantified in
Fig.~\ref{fig: Performance fct r},
where we show how the political,
health, and economic cost decreases with increasing test rate.

\subsubsection{Time delay to detect catastrophic growth rates}

After a reboot it is likely that the growth rate $k^{\,}_{1}$ jumps back to
positive values, as we have always assumed so far. The time it takes
until one can distinguish a genuine growth from intrinsic fluctuations
due to the finite number of sample people depends on the growth rate
$k^{\,}_{1}$, see Eq.~(\ref{Deltat1}).

\begin{figure}[t]
\begin{center}
\includegraphics[width=0.95\linewidth]{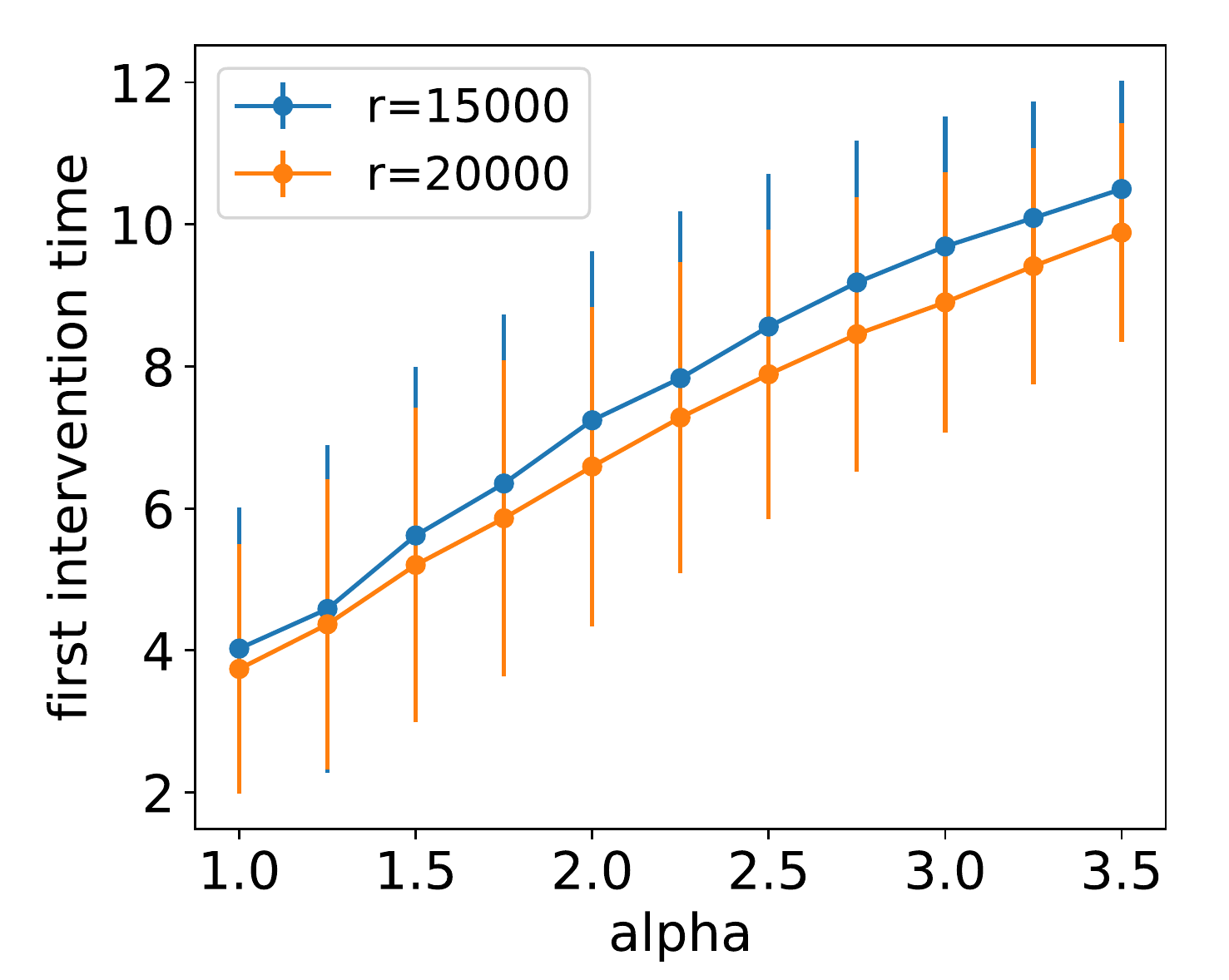}
\end{center}
\caption{
Time after which a significant positive growth rate is confirmed
in the worst case scenario for which the growth rate
jumps to $k^{\,}_{1}=0.23$ after reboot.
An intervention will be triggered in 3-4 days,
since in the case that such a strong growth must be suspected,
one should apply a small confidence parameter $\alpha\approx 1$.
Results are shown for $r=15'000$ and $r=20'000$ tests a day.
The circles are the mean values,
the vertical lines indicate the standard deviations
for the first intervention time.
        }
\label{fig: Delta time 1}
\end{figure}

In the worst case where the reboot brings back the unmitigated value
$k^{\,}_{0}$, one will know within 3-4 days
with reasonable confidence that the growth
rate is well above zero. This is shown in Fig.~\ref{fig: Delta time 1}.
In such a catastrophic situation, an early intervention can be taken,
while the number of infections has at most tripled at worst.
Note that this reaction time is 4-5 times faster than without
random testing.

\section{Regionally refined reboot  and mitigation strategies}
\label{sec: Regionally refined reboot  and mitigation strategies}

We have argued that the minimal testing rate $r^{\,}_{\mathrm{min}}$
(\ref{testfrequency}) is sufficient to obtain statistical
information on the growth rate $k$ as applied to Switzerland as a
whole. This assumes tacitly that the simple growth equation
(\ref{eq: growthequation}) describes the dynamics of infections in the
whole country well.  That this is not necessarily a good description
can be conjectured from recent data on the current rates with which
numbers of confirmed infections in the various cantons
(states of Switzerland) evolve.
These data indeed show a very significant spread by
nearly a factor of four suggesting that a spatially resolved approach
is preferable, if possible. 

If the testing capacity is limited by rates of order
$r^{\,}_{\mathrm{min}}$, the approach can still be used. But caution
should be taken to account for spatial fluctuations corresponding to
hot spots. One should preferentially test in areas that are likely to
show the largest local growth rates so as not to miss locally
super-critical growth rates by averaging over the entire country. If
however, higher testing frequencies become available, new and better
options come into play.
  
\subsection{Partitioning the country for statistical analysis}

Valuable information can be gained by analyzing the test data not only
for Switzerland as a whole, but by distinguishing different regions.

It might even prove useful not to lift restrictions homogeneously
throughout the country, but instead to vary the set of restrictions to
be released, or to adapt their rigor.  By way of example, consider
that after the spring vacation school starts in different weeks in
different cantons. This regional
difference could be exploited to probe the relative effect of
re-opening schools on the
local growth rates $k$.  However, obviously, it might prove
politically difficult to go beyond such ``naturally'' occurring
differences, as it is a complex matter to decide what
region releases which measures first.  A further issue is that the
effects might be unclear at the borders between regions with different
restrictions. There may also be complications with commuters that
cross regional borders. Finally, there may be undesired behavioral
effects, if regionally varying measures are declared as an
``experiment''. Such issues demand careful consideration if regionally
varying policies are applied.

Even if policy measures should eventually not be taken in a
region-specific manner, it is very useful to study a regionally
refined model of epidemic dynamics. Indeed a host of literature exists
that studies epidemiological models on lattices and analyzes the
spatial heterogeneities.%
~\cite{Grassberger83,Cardy85}
In certain circumstances those have been argued to become even extremely
strong.%
~\cite{Vojta09}
In the present paper, we will content
ourselves with a few general remarks concerning such refinements. We
reserve a more thorough study of regionally refined testing and
mitigation strategies to a later publication.

Let us thus group the population of Switzerland into $G$ sets. The most
natural clustering is according to the place where people live, cities
or counties.%
~\footnote{
One might also consider other distinguishing
characteristics of groups (age or commuting habits, etc.), but we will
not do so here, since it is not clear whether the increased
complexity of the model can be exploited to reach an improved data
analysis. In fact we expect that the number of fitting parameters
will very quickly become too large by making such further
distinctions.
          }
The more we partition the country, the more spatially refined
the acquired data will be,
and the better tailored mitigation strategies could potentially 
become. However, this comes at a price.
Namely, for a limited national testing rate $r^{\,}_{\mathrm{tot}}$,
an increased partitioning means that the statistical uncertainty to
measure local growth rates in each region will increase.

The minimal test rate $r^{\,}_{\mathrm{min}}$  that we estimated
on the right-hand side of
Eq.\ (\ref{testfrequency})
still holds, but now for each region, which can only test at a rate
$r=r^{\,}_{\mathrm{tot}}/G$. To refine Switzerland $G$ regions
we thus have the constraint that the total testing capacity exceeds
\be
r^{\,}_{\mathrm{tot}}\geq
G\,r^{\,}_{\mathrm{min}}\equiv
G\,\frac{(4\,\alpha)^{2}}{(\ln {4})^{3}}\frac{k^{\,}_{1}}{i^{*}}.
\ee
On the other hand, if testing at a high daily rate
$r^{\,}_{\mathrm{tot}}$ becomes available, nothing should stop one to
refine the statistical analysis to
$G\approx r^{\,}_{\mathrm{tot}}/r^{\,}_{\mathrm{min}}$
to make the best use of available data.

\subsection{Spatially resolved growth model}

Each of the population groups $m\in\{1,\cdots,G\}$ is assumed to have roughly
the same size, containing
\be
N^{\,}_{m}\approx
\frac{N^{\,}_{\mathrm{CH}}}{G} 
\ee
people, $U^{\,}_{m}$ of whom are currently infected, but yet undetected.
The spreading of infections is again assumed to follow a linear
growth equation
(where we neglect influx from across the borders from the outset)
\be
\label{growthequation bis}
\left(\frac{\mathrm{d}U^{\,}_{m}}{\mathrm{d}t}\right)(t)=
\sum_{n=1}^{G}
K^{\,}_{mn}(t)\,
U^{\,}_{n}(t),
\qquad
m=1,\cdots,G.
\ee
Here, the growth kernel $K(t)$ is a $G\times G$ matrix  with matrix elements
$K^{\,}_{mn}(t)$. The matrix $K(t)$ has $G$ (complex valued)
eigenvalues $\lambda^{\,}_{n}$, $n=1,\cdots,G$. 
The largest growth rate is given by
\be
\kappa(t)\equiv
\max\limits^{\,}_{1\leq n\leq G}\,
\left\{\mathrm{Re}\,\lambda^{\,}_{n}(t)\right\}.
\ee
For the sake of stability criteria, $\kappa(t)$ now essentially takes the
role of $k(t)$ in the model with a single region, $G=1$.  We note that the
number of infections grows exponentially if $\kappa(t)>0$, and decreases
if $\kappa(t)<0$.

As in the case of a single region, we assume $K(t)$ to be piece wise
constant in time, and to change only upon taking policy interventions.

In the simplest approximation, one assumes no contact between geographically
distinct groups, that is, the off-diagonal matrix elements are set to zero
[$K^{\,}_{m\neq n}(t)=0$] and the eigenvalues become equal to elements of the
diagonal: $k^{\,}_{m}(t)\equiv K^{\,}_{mm}(t)$.
As current  cantonal data suggests, the local growth
rate $k^{\,}_{m}(t)$ depends on the region, and thus
$k^{\,}_{m}(t)\neq k^{\,}_{n}(t)$.  It
is natural to expect that $k^{\,}_{m}(t)$ correlates with the population
density, the fraction of the population that commutes, the age
distribution, etc. 

If on top of the heterogeneity of growth rates one adds finite but
weak inter-regional couplings $K^{\,}_{m\neq n}(t)>0$ (mostly between
nearest neighbor regions), one may still expect the eigenvectors of
$K(t)$ to be rather localized (a phenomenon well known as Anderson
localization~\cite{Anderson58} in the context of waves propagating in
strongly disordered media). By this, one means that the
eigenvectors have a lot of weight on few regions only, and little
weight everywhere else.
That such a phenomenon might occur in the growth pattern of real
epidemics is suggested by the significant regional differences in
growth rates that we have mentioned above. In such a
situation it would seem preferable to adapt restrictive measures
to localized regions with strong overlap on unstable eigenvectors
of $K(t)$, while minimizing their socio-economic impact in other
regions with lower $k^{\,}_{m}(t)$.

\subsection{Mitigation strategies with regionally refined analysis}
 
As mentioned above, in the case with several distinct regions, $G>1$,
an intervention becomes necessary when the largest eigenvalue
$\kappa(t)$ of $K(t)$ crosses an upper or a lower threshold (with a
level of confidence $\alpha$ again to be specified).  If the
associated eigenvector is delocalized over all regions, one will most
likely respond with a global policy measure.  However, it may as well
happen that the eigenvector corresponding to $\kappa(t)$ is
well-localized. In this case one can distinguish two strategies for
intervention:

\begin{enumerate}
\item[(a)] \textbf{Global strategy}
One always applies a single policy change to the whole country.
This is politically simple to implement, but might incur unnecessary economic
cost in regions that are not currently unstable.
\item[(b)] \textbf{Local strategy}
One applies a policy change only in regions which have significant
weight on the unstable eigenvectors. This means that one only adjusts
the corresponding diagonal matrix elements of $K(t)$ and those off-diagonals
that share an index with the unstable region.
\end{enumerate}

Likewise, regions that have $i^{\,}_{m}<i^{*}$ and have negligible
overlap with eigenvectors whose eigenvalues are above $\kappa^{\,}_{-}$,
could relax some restrictions before others do.

Fitting test data to a regionally refined model will allow us to
estimate the off-diagonal terms $K^{\,}_{mn}(t)$, which are so far poorly
characterized parameters. However, the $K^{\,}_{mn}(t)$ contain valuable
information. For instance, if a hot spot emerges [that is, a region
overlapping strongly with a localized eigenvector with positive
$\mathrm{Re}\,\lambda^{\,}_{n}(t)$], this part of the matrix will inform which
connections are the most likely to infect neighboring regions. They
can then be addressed by appropriate policy measures and will be
monitored subsequently, with the aim to contain the hot spot and keep
it well localized.

This model allows us to calculate again economic, political,
and health impact of various strategies. It is important to assess how
the global and the local strategy perform in comparison. Obviously
this will depend on the variability between the local growth rates
$k^{\,}_{m}(t)$, which is currently not well known, but will become a measurable
quantity in the future. At that point one will be able to decide
whether to select the politically simpler route (a) or the
heterogeneous route (b) which is likely to be economically favorable.

We are currently engaged in developing an analysis tool to quickly
process test data for multi-region modeling. We are developing and
assessing intervention strategies with the perspective of running it
daily with the best available current data and knowledge. 

\section{Summary and conclusion}
\label{sec: Summary and conclusion}

We have analyzed a feedback and control model for managing a pandemic
such as that caused by COVID-19. The crucial output parameters are the
infection growth rates in the general population and spatially
localized sub-populations. When planning for an upcoming reboot of the
economy, it is essential to assess and mitigate the risks of relaxing
some of the restrictions that have brought the COVID-19 epidemic under
control. In particular, the policy strategy chosen must suppress a
potential second exponential wave when the economy is rebooted, and so
avoid a perpetual stop-and-go oscillation between relaxation and
lockdown. Feedback and control models are designed with precisely
this goal in mind.

Having random testing in place, the risk of a second wave can be kept
to a minimum. Additional testing capacity of
$r=15'000\,\mathrm{day}^{-1}$
tests (on top of the current tests for medical purposes)
carried out with randomly selected people would allow us to
follow the course of the pandemics almost in real time, without huge
time delays, and without the danger of increasing the number of
currently infected people by more than a factor of four, if our intervention
strategy is followed. We emphasize that our estimate of $r$ is conservative.
If the manageable fraction of infected people is higher than what
we assumed in Eq.\ (\ref{eq: def ic}),
namely of order $i^{\,}_{\mathrm{c}}\approx 0.01$
as the estimates of Ref.~\cite{FergusonReport13} suggest,
the required testing rate  decreases by a factor $3-4$ to a mere
$r=4'000-5'000\,\mathrm{day}^{-1}$.

If testing rates $r$ significantly higher than $r^{\,}_{\mathrm{min}}$
become available, a regionally refined analysis of the growth dynamics
can be carried out, with $G\approx r/r^{\,}_{\mathrm{min}}$ regions
that can be distinguished.
 
In the worst case scenario, where releasing certain measures
immediately make the country jump back to the unmitigated growth rate
of $k^{\,}_{0}=0.23\,\mathrm{day}^{-1}$, random testing would detect this
within 3-4
days from the change coming into effect. This is in stark
contrast to the nearly 14 days of delay required for
symptomatic individuals to emerge in statistically significant numbers.
After such a time delay a huge increase (by a factor of order 20)
of infection numbers may have already occurred, which would be catastrophic.
Daily random testing safely prevents this. Thereby
the significant reduction of the time delay is absolutely crucial.
Note that without daily polling of infection numbers and without
knowledge about the quantitative effect of restriction measures, a
reboot of the economy could not be risked before the number of
infections has been suppressed by at least a factor of 10-20 below the
current level. Given the limits of suppression rates that can be
achieved without most draconic lockdown measures, this will
require a very long time and thus translates into an enormous economic
cost.  In contrast, daily polling will allow us to carefully
reboot the economy and adjust restrictive measures, while closely
monitoring their effect. Since the reaction times are so much shorter,
one can safely start an attempted reboot already at infection numbers
corresponding roughly to the status quo.

At some point one might consider the option to start releasing 
different sets of restrictions in different regions,
with the aim to learn faster about
their respective effects and thus to optimize response strategies
in subsequent steps. 

\section{Acknowledgments}
\label{sec: Acknowledgments}

We are grateful to
Giulia Brunelli, Klaus M\"ueller, Emma Slack, Thomas Van Boeckel,
and Li Zhiyuan for helpful discussions,
and the ERC HERO project 810451 for supporting GA.  

\appendix

\section{Assessment of contact tracing as a means to control the pandemics}

Let us briefly discuss the strategy of so-called contact tracing as a means
to contain the pandemics, as has been discussed in the literature%
~\cite{Hollingsworth20}. We argue that contact tracing is a helpful tool to
suppress transmission rates, but is susceptible to fail when no other
method of control is used.

Contact tracing means that once an infected person is detected, people
in their environment (i.e., known personal contacts, and those
identified using mobile-phone based Apps etc) are notified and tested,
and quarantined if detected positive. As a complementary measure to
push down the transmission rate, it is definitely useful,
and it represents a relatively low
cost and targeted measure, since the probability to detect infected people
is high. However, as a sole measure to contain a pandemic
contact tracing is impractical
(especially at the current high numbers of infected people)
and even hazardous.

The reason is as follows. It is believed that a considerable fraction
$f^{\,}_{\mathrm{asym}}$ of infected
people show only weak or no symptoms, so that they would not get tested
under the present testing regime. The value of
$f^{\,}_{\mathrm{asym}}$ is not well known, but it might be rather high (30\%
or even much higher). Such asymptomatic people will go undetected,
if they have not been in contact with a person displaying symptoms.
If on average they infect $R$ people while being infectious, and if
$R\,f^{\,}_{\mathrm{asym}}>1$, there will be an exponential avalanche of undetected
cases. They will produce an exponentially growing number of detectable
and medically serious cases. The contact tracing of those
(upward in the infection tree) is tedious, and 
cannot fully eliminate the danger of such an avalanche.

Contact tracing as a main strategy thus only becomes viable once the value of
$f^{\,}_{\mathrm{asym}}$ is well established,  and
one is certain to be able to control the value of $R$ such that
$R\,f^{\,}_{\mathrm{asym}}<1$.

\section{Algorithm  to simulate mitigation of reboot}
\label{app:algorithm}

\subsection{Definitions}

\begin{itemize}
	\item $t=1,2,\cdots$:
	Time in days (integer).
	
	\item $n^{\,}_{\mathrm{int}}$:
	Number of interventions (including the reboot at $t=1$).
		
	\item $t^{\,}_{\mathrm{int}}(j)$:
	First day on which the $j$'th rate $k^{\,}_{j}$ applies.
	On day $t^{\,}_{\mathrm{int}}(1)\equiv 1$
	the initial reboot step is taken.
	
	\item $\Delta t(j)=t^{\,}_{\mathrm{int}}(j+1)-t^{\,}_{\mathrm{int}}(j)$:
	Time span between interventions $j$ and $j+1$. 
	
	\item $t^{\,}_{\mathrm{first}}$:
	First day on which the current rate $k=k(t)$ is applied.
	
	\item $i(t)$:
	Fraction of infected people on day $t$.
	
	\item $k(t)$:
	Growth rate on day $t$. 
	
	\item $r$:
	Number of tests per day.
	
	\item $C^{\,}_{\mathrm{H}}$:
	Health cost.
	
	\item $C^{\,}_{\mathrm{E}}$: Economic cost.
	
	\item  $k^{\,}_{\mathrm{min}}=0.005$:
	Minimal growth rate targeted.
	
	\item $i^{\,}_{\mathrm{low}}=0.2$:
	Lower threshold for $i/i^{*}$.
	If $i/i^{*} < i^{\,}_{\mathrm{low}}$,
        a relaxing intervention is made,
        irrespective of the estimate of $k$.
        	
	\item $i^{\,}_{\mathrm{high}}=3$:
	Upper threshold for $i/i^{*}$.
	If $i/i^{*} > i^{\,}_{\mathrm{high}}$,
	an intervention is made even if $k$ is still smaller than
	$\alpha\,\delta k$.
	
	\item $k^{\,}_{\mathrm{low}}=-0.1$:
	Minimal possible decreasing rate considered.
	
	\item $k^{\,}_{\mathrm{high}}=0.23$:
	Maximal possible increasing rate considered.
	
	\item $T^{\,}_{\mathrm{min}}=3$:
	Minimal time to wait since the last intervention,
	for interventions based on the level of $i(t)$.
	
	\item $b$:
	Parameter defining the possible range of changes $\Delta k$
        due to measures taken after estimating $k$.
	$|\Delta k/k^{\,}_{\mathrm{est}}|\in[b, 1/b]$.
        Usually we set $b=0.5$.
	
	\item $\alpha$: Confidence parameter. 
	
	\item $N(t)$:
	Cardinality of random sample of infected people on day $t$. The number
	$N(t)$ is obtained by sampling from a Gaussian distribution of mean
        $i(t)\,r$ and standard deviation
	$\sqrt{i(t)\,r}$ and rounding the obtained real number to the next
	non-negative integer.
	
\end{itemize}

\subsection{Initialization}
\begin{itemize}
	
	\item $t^{\,}_{\mathrm{first}}=t^{\,}_{\mathrm{int}}(1)=1$.
	
	\item $n^{\,}_{\mathrm{int}}=1$.
	
	\item $C^{\,}_{\mathrm{H}}=1$.
	
	\item $C^{\,}_{\mathrm{E}}=0$.
	
	\item $k(1)=k^{\,}_{1}= 0.1$. (Initial growth rate)
	
	\item $i(1)= i^{*}$. Common choice $i^{*}=i^{\,}_{\mathrm{c}}/4=0.0007$.
	
	\item Draw $N(1)$.
	
	\item $k(2)= k(1)$. (No intervention at the end of day 1)
	
	\item Set $t=2$.
	
\end{itemize}

\subsection{Daily routine for day $t$}

\noindent
Define $i(t)=i(t-1)\,e^{k\,(t-1)}$,

\noindent
Define $C^{\,}_{\mathrm{H}}=\max\{C^{\,}_{\mathrm{H}},i(t)/i^{*}\}$,

\noindent
Define $C^{\,}_{\mathrm{E}}=C^{\,}_{\mathrm{E}}-k(t)$.

\noindent
Draw $N(t)$.

\noindent
Determine what will be $k(t+1)$, by assessing whether or not to intervene: 

\noindent
\textbf{If} $t= t^{\,}_{\mathrm{first}}$,
\textbf{then} $k(t+1)=k(t)$. (No intervention)

\noindent
\textbf{Else} Distinguish three intervention cases:
\begin{enumerate}
	\item
	\textbf{If}
	$i(t)/i^{*}<i^{\,}_{\mathrm{low}}$
	and
	$t-t^{\,}_{\mathrm{first}}\geq T^{\,}_{\mathrm{min}}$,
	\textbf{then}\\
	$k(t+1)=\min\{k(t)+x\,k^{\,}_{1}, k^{\,}_{\mathrm{high}}\}$\\
	with $x=\mathrm{Unif[0,1]}$.
	\item
	\textbf{ElseIf}
	$i(t)/i^{*}>i^{\,}_{\mathrm{high}}$
	and
	$t-t^{\,}_{\mathrm{first}}\geq T^{\,}_{\mathrm{min}}$, \textbf{then}\\
	$k(t+1)=\max\{k(t)-(1+x)/2\,k^{\,}_{\mathrm{high}},k^{\,}_{\mathrm{low}}\}$\\
	with $x=\mathrm{Unif[0,1]}$.
	
	\item
	\textbf{ElseIf} $i^{\,}_{\mathrm{low}}<i(t)/i^{*}<i^{\,}_{\mathrm{high}}$,
	\textbf{then}
	\begin{itemize}
		\item
		set ${\Delta t}\equiv t-t^{\,}_{\mathrm{first}}+1$
		\item
		Compute
		$k^{\,}_{\mathrm{est}}(t^{\,}_{\mathrm{first}},
		{\Delta t})$,
		and
		$\delta k^{\,}_{\mathrm{est}}(t^{\,}_{\mathrm{first}},{\Delta t})$
		using Sec.\ \ref{subsec: Estimate of k(t,{Delta t})}.
		
		\textbf{If}  
		$
		|k^{\,}_{\mathrm{est}}|>k^{\,}_{\mathrm{min}}
		$\\
		\textbf{AND}\\
		$
		\left[
		k^{\,}_{\mathrm{est}}>\alpha\,\delta k^{\,}_{\mathrm{est}}
		\hbox{ \textbf{OR} }
		k^{\,}_{\mathrm{est}}<
		-
		\alpha\,\delta k^{\,}_{\mathrm{est}}
		\right]
		$,\\
		set\\
		$k(t+1)=k(t)-x\,k^{\,}_{\mathrm{est}}$\\
		with $x=\mathrm{Unif}[b,1/b]$.
		
		If $k(t+1)>k^{\,}_{\mathrm{high}}$,
                put $k(t+1)=k^{\,}_{\mathrm{high}}$.\\
		
		If $k(t+1)< k^{\,}_{\mathrm{low}}$,
                put $k(t+1)=k^{\,}_{\mathrm{low}}$.
		
	\end{itemize}  
\item \textbf{Else} $k(t+1)= k(t)$
\end{enumerate}
\noindent
$t=t+1$.

\noindent
\textbf{If} an intervention was taken above:
\begin{itemize}
	\item Put $n^{\,}_{\mathrm{int}}=n^{\,}_{\mathrm{int}}+1$.
	\item Define $t^{\,}_{\mathrm{int}}(n^{\,}_{\mathrm{int}})=t+1$. 
	\item Define $\Delta t({n^{\,}_{\mathrm{int}}-1})=
	  t^{\,}_{\mathrm{int}}(n^{\,}_{\mathrm{int}})
          -
          t^{\,}_{\mathrm{int}}(n^{\,}_{\mathrm{int}}-1)$.
	\item Set $t^{\,}_{\mathrm{first}}= t+1$.
\end{itemize}

\noindent
\textbf{If}
$|k^{\,}_{\mathrm{est}}|<k^{\,}_{\mathrm{min}}$ 
\textbf{AND} $k(t)<k^{\,}_{\mathrm{min}}$
\textbf{AND} $t-t^{\,}_{\mathrm{first}}>10$, \textbf{then} 
EXIT. 

\noindent
\textbf{Else} Return to daily routine for next day.

\subsection{Estimate of $k(t,{\Delta t})$}
\label{subsec: Estimate of k(t,{Delta t})}

\noindent
Computing
$k^{\,}_{\mathrm{est}}(t^{\,}_{\mathrm{first}},{\Delta t})$
and
$\delta k^{\,}_{\mathrm{est}}(t^{\,}_{\mathrm{first}},{\Delta t})$:

\noindent
\textbf{If} ${\Delta t}$ is even:

\noindent
Define 

\noindent
$N^{\,}_{1}=
\sum\limits_{m=0}^{{\Delta t}/2-1}N(t^{\,}_{\mathrm{first}}+m)$,

\noindent
$N^{\,}_{2}=
\sum\limits_{m=0}^{{\Delta t}/2-1} N(t^{\,}_{\mathrm{first}}+{\Delta t}/2+m)$. 

\begin{itemize}
	\item
	\textbf{If}
        $N^{\,}_{1}\,N^{\,}_{2} >0$,
	\textbf{then}\\
	$k^{\,}_{\mathrm{est}}=
	\frac{2}{{\Delta t}}\,
	\ln\left(\frac{N^{\,}_{2}}{N^{\,}_{1}}\right)$,\\
	\noindent
	$\delta k^{\,}_{\mathrm{est}}=
	\frac{2}{{\Delta t}}\sqrt{\frac{1}{N^{\,}_{2}}+\frac{1}{N^{\,}_{1}}}$.
	
	\item
	\textbf{Else}
	return\\
	$k^{\,}_{\mathrm{est}} =0$,\\
	$\delta k^{\,}_{\mathrm{est}}=1000$.\\
	
\end{itemize}

\textbf{If}  ${\Delta t}$ is odd:\\

\noindent
Define\\

\noindent
$
N^{\prime}_{1}=
\sum\limits_{m=0}^{({\Delta t}-1)/2-1}
N(t^{\,}_{\mathrm{first}}+m)
$,\\

\noindent
$
N^{\,}_{m}=N(t^{\,}_{\mathrm{first}}+({\Delta t}-1)/2)
$,\\

\noindent
$
N^{\prime}_{2}=
\sum\limits_{m=0}^{({\Delta t}-1)/2-1}
N(t^{\,}_{\mathrm{first}}+({\Delta t}+1)/2+m)
$,\\ 

\noindent
$
N^{\,}_{1}=
N^{\prime}_{1}+N^{\,}_{m}
$,\\

\noindent
$
N^{\,}_{2}=
N^{\prime}_{2}+N^{\,}_{m}
$.\\

\begin{itemize}
	\item  
	\textbf{If} $N^{\,}_{1}\,N^{\,}_{2}>0$, \textbf{then}\\
	\noindent
	$k^{\,}_{\mathrm{est}}=
	\frac{2}{({\Delta t}-1)}\,
	\ln\left(
	\frac{N^{\prime}_{2}+N^{\,}_{m}}{N^{\prime}_{1}+N^{\,}_{m}}
	\right)$,\\
	\noindent
	$\delta k^{\,}_{\mathrm{est}}=
	\frac{2}{({\Delta t}-1)}\,
	\sqrt{\frac{N^{\prime}_{2}}{N^{2}_{2}}
		+
		\frac{N^{\prime}_{1}}{N^{2}_{1}}
		+
		N^{\,}_{m}\,
		\left(
		\frac{1}{N^{\,}_{2}}-\frac{1}{N^{\,}_{1}}
		\right)^{2}}$.
	\item
	\textbf{Else} return\\
	\noindent
	$k^{\,}_{\mathrm{est}} =0$,\\
	$\delta k^{\,}_{\mathrm{est}}=1000$.
\end{itemize}

\subsection{Observables}

Time to first intervention:
$\Delta t(1)$

Health cost:
$C^{\,}_{\mathrm{H}}$

Political cost:
$n^{\,}_{\mathrm{int}}$

Economic cost
$C^{\,}_{\mathrm{E}}$

\end{document}